\documentclass[twocolumn]{aastex6}

\pdfoutput=1

\newcommand{\luna}{{\tt LUNA}}
\newcommand{\multi}{{\sc MultiNest}}
\newcommand{\cofiam}{{\tt CoFiAM}}
\newcommand{\kepler}{\emph{Kepler}}
\newcommand{\grand}{grand light curve} 
\newcommand{\NEA}{\href{http://exoplanetarchive.ipac.caltech.edu/index.html}{the NASA Exoplanet Archive}}

\usepackage{amsmath}
\usepackage{amsfonts}
\usepackage{amssymb}
\usepackage{wasysym}
\usepackage{comment}

\begin{document}


\title{HEK VI: On the Dearth of Galilean Analogs in \textit{Kepler}, \\
and the Exomoon Candidate Kepler-1625b I}


\author{
A.~Teachey\altaffilmark{1},
D.~M.~Kipping\altaffilmark{1} \&
A.~R.~Schmitt\altaffilmark{2}
}

\affil{ateachey@astro.columbia.edu}

\altaffiltext{1}{Department of Astronomy, Columbia University, 550 W 120th St., New York, NY 10027}

\altaffiltext{2}{Citizen Science}

\begin{abstract}

Exomoons represent an outstanding challenge in modern astronomy, with the potential
to provide rich insights into planet formation theory and habitability. In this work, we stack the phase-folded transits of 284 viable moon hosting \textit{Kepler} planetary candidates, in order to search for satellites. These planets range from Earth-to-Jupiter sized and from ${\sim}$0.1-to-1.0\,AU in separation - so-called ``warm'' planets. Our data processing includes two-pass harmonic detrending, transit timing variations, model selection and careful data quality vetting to produce a grand light curve with a r.m.s. of 5.1\,ppm. We find that the occurrence rate of Galilean-analog moon systems for planets orbiting between ${\sim}$0.1 and 1.0\,AU can be constrained to be $\eta<0.38$ to 95\% confidence for the 284 KOIs considered, with a 68.3\% confidence interval of $\eta=0.16_{-0.10}^{+0.13}$. A single-moon model of variable size and separation locates a slight preference for a population of short-period moons with radii ${\sim}0.5$\,$R_{\oplus}$ orbiting at 5-10 planetary radii. However, we stress that the low Bayes factor of just 2 in this region means it should be treated as no more than a hint at this time. Splitting our data into various physically-motivated subsets reveals no strong signal. The dearth of Galilean-analogs around warm planets places the first strong constraint on exomoon formation models to date. Finally, we report evidence for an exomoon candidate Kepler-1625b I, which we briefly describe ahead of scheduled observations of the target with the Hubble Space Telescope.

\end{abstract}




\section{INTRODUCTION}
\label{sec:intro}


Moons present unique scientific opportunities. In our Solar System, they offer
clues to the mechanisms driving early and late planet formation, and several of
them are thought to be promising targets in the search for life, as several are rich
in volatiles \citep[e.g.][]{squyres:1983, hansen:2006} and possess internal heating mechanisms \citep[e.g.][]{morabito:1979, hansen:2005, sparks:2016}. 
The moons of our Solar System also demonstrate the great variety of geological features that may be found on other terrestrial worlds. 

In this new era of exoplanetary science it stands to reason that moons in
extrasolar systems, so-called exomoons, should tell us a great deal about the
commonality of the processes that shaped our Solar System and may yield just as
many surprises as their host planets before them. Just as the study of
exoplanets has complicated our picture of planetary formation by revealing
(for example) the existence of Hot Jupiters \citep{mayor:1995} -- worlds
without Solar System analogs -- so too might moons show us what else is
possible and uproot conventional thinking about satellite formation mechanisms.

Galilean-sized moons ($\sim0.2$-$0.4$\,$R_{\oplus}$) are generally thought to
be able to form in a variety of ways. For the regular satellites of Jupiter,
the Galilean moons are thought to have condensed out of a circumplanetary disk,
akin to planet formation within a protoplanetary disk \citep{canup:2002}. This
process is expected to limit regular satellites to a cumulative mass of
$\mathcal{O}[10^{-4}]$ that of the primary \citep{canup:2006}. Higher
mass-ratio moons, such as the Earth's Moon, are evidently viable too and may
form from catastrophic collisions in the first few hundred million years of the
solar system, coalescing from that collision's debris \citep[e.g.][]{ida:1997}. Finally,
retrograde Triton is hypothesized to have originated from a capture event via a
binary exchange mechanism \citep{agnor:2006}. Put together, Galilean-sized
satellites appear to have formed via at least three independent pathways within
the Solar System, and their existence around exoplanets can therefore be
reasonably hypothesized.

Galilean-sized exomoons are challenging to detect using the transit
method\footnote{Note that the transit method is by no means the only method
sensitive to exomoons; microlensing, for example, is another promising avenue \citep{bennett:2014}.} for a number of reasons.
First, the transit of a $0.2$-$0.4$\,$R_{\oplus}$ moon across a Sun-like star
results in a depth of $3$-$13$\,ppm, below the typical sensitivity
achievable with \textit{Kepler} \citep{christiansen:2012}. Second, the moon
signal will almost certainly be found at each epoch in a different location
with respect to the host planet, sometimes occurring before the transit,
sometimes after, and at a different projected distance from the planet
\citep{luna:2011}. Third, multiple moons around a single planet may wash out
any transit timing \citep{sartoretti:1999} or duration variations
\citep{TDVa:2009,TDVb:2009}. The three-body motion combined with the potential
for overlapping disks (syzygies), makes accurate modeling of exomoon transits
computationally demanding \citep{luna:2011}. Fourth, the long-period nature of
plausible moon hosts means that relatively few transits are usually available.
In this regime, a planet+moon light curve model has sufficient flexibility to
almost always provide a superior fit to the limited data \citep{bash:2013},
thereby necessitating rigorous Bayesian approaches to model selection. Despite
these challenges, the ``Hunt for Exomooons with Kepler'' (HEK) project has
performed Bayesian photodynamical fits of ${\sim}60$ exoplanets to date
\citep{HEK1,HEK2,HEK3,HEK4,HEK5}, with no unambiguous detections and limits
typically hovering around an Earth-mass.

In this work, our project pursues a different approach to searching for
evidence of moons which focuses on seeking a population of moons around a
population of planets. Rather than pursuing individual limits which are then
combined to constrain the population, we here approach the problem from a
broad statistical perspective in order to directly measure the occurrence rate
of moons, $\eta_{\leftmoon}$. Resulting from this analysis, we identify a
single exomoon candidate, Kepler-1625b I. We briefly describe its detection and vetting
ahead of scheduled observations of the planet with the Hubble Space Telescope in October 2017.

\section{STACKING EXOMOONS}
\label{sec:concept}

\subsection{Phase-Folding}

The work presented here aims to exploit the power of stacking in order
to search for exomoons. Stacking is a familiar technique to those
studying exoplanet transits, who typically phase-fold a light curve
upon the period. For a linear ephemeris, the transits align leading to
a coherent signal. It is important to stress that this act does not improve
the signal-to-noise ratio (SNR). The amount of data before and after stacking is
the same, with the only difference being that stacking assumes the ephemeris
of the planet is known to infinite precision. Modeling the full unstacked
light curve with a model conditioned upon the same ephemeris assumption would
result in identical posteriors and thus no improvement is achieved for the
inference itself. Despite this, stacking is attractive because the signal's
coherence means that full light curve modeling is unnecessary in the context
of signal detection. Specifically, one may simply evaluate the weighted mean
centered around the pivot point of the fold to achieve a detection, which is why
the popular Box Least Squares (BLS) algorithm \citep{BLS:2002} is a
computationally efficient yet sensitive tool in conventional transit detection.

Stacking light curves in pursuit of exomoons is complicated by the fact that
simply phase-folding light curves upon a linear ephemeris will lead to
the moon appearing at different phases in each epoch. Despite the fact that the moon is not perfectly coherent, it is constrained to lie within a fraction of the 
Hill sphere radius \citep{barnes:2002} and this imparts some quasi-coherent 
properties into the phase-folded light curve. \citet{simon:2012} were the first to describe
this possibility, where they argued that this quasi-coherence will lead to
an increase in the photometric scatter in the temporal region surrounding
the planetary event - an effect they dubbed ``scatter peak''. A similar
idea is discussed in \citet{heller:2014}, who instead considered looking
for a slight photometric decrease in this temporal region. By considering
the probability density of the moon's sky-projected position, an effect dubbed 
the ``orbital sampling effect'' (OSE), \citet{heller:2014}
derives a simple formula for predicting phase-folded light curve
shape in the presence of moons, enabling a simple approach to seeking
exomoon shadows.

As with the case of a planetary transit, or indeed any kind of stacking,
this approach does not boost SNR in any way. The data volume and quality
are the same before and after the stacking. However, unlike the planet
case, the shape of the phase-folded moon signal represents a washed-out
depiction of the individual signals. Accordingly, the subtle individual
variations in durations, positions and shapes are lost, meaning that
stacking imposes a fundamental loss of information content, and
therefore sensitivity.

A similar situation occurs when observing planetary transits with
long exposures, such as the 30\,minute cadence (LC) mode of \textit{Kepler},
causing the shape of the transit to be slightly washed-out, thereby degrading
the information content. It is for this reason that short-cadence (SC)
\textit{Kepler} data provides tighter constraints on transit times, despite
the fact that both see the same SNR transit depth (e.g. see \citealt{HEK2} and
\citealt{HEK4}).

Accordingly, searching for exomoons in phase-folded light curves will always
be less sensitive than full photodynamic fits -- although precisely how much
has not been formally evaluated and would be sensitive to the specific
planet-moon parameters. 
Despite this, phase-folding is attractive for its simplicity and as a pure
detection tool -- analogous to BLS for planet hunting.

One particularly attractive feature of phase-folding exomoons is that a
sizable fraction of the quasi-coherent signal appears exterior to the
planetary transit. Assuming the planetary transits are well-aligned and
the duration is well-known, one may simply crop the planetary transit
leaving behind a pure exomoon signal. This greatly simplifies the analysis,
since the planet properties are not covariant with this signal\footnote{Phase variations are not included in our model and thus cannot be covariant, nor should such phase variations persist given our detrending algorithm will have largely removed them.}. For these
reasons, we identify this out-of-transit phase-folded moon signal as the
target signature in this work.

\subsection{Planet-Stacking}

Unlike a simple planetary phase-fold, the quasi-coherent nature of the
phase-folded moon light curve means that a large number of transits
are needed to produce a predictable signal. At its core, the phase-folded
moon signal depends upon the law of averages and thus co-adding relatively
few transits can lead to a phase-folded moon signal which is highly irregular
and erratic. Without a characteristic and predictable shape, it is very
difficult to convincingly argue the signal is genuinely a moon, rather than
some peculiarity of the data in those limited co-added events. Indeed,
\citet{heller:2014} argue that at least a dozen events are generally needed,
such that $N\gg1$ and the averaging effect can become noticeable.

Unfortunately, \textit{Kepler}'s primary mission lifetime of 4.35 years means
that the long-period planets, where moons are most \textit{a priori} expected to be
viable, were only observed to transit a few times. Only in a small number of
cases are there \textit{Kepler} planets for which their period is long enough
such that moons are dynamically stable for Gyr \textit{and} we possess
$N\gg1$ transit events within the 4.35 years of \textit{Kepler} observations.
This point seemingly excludes phase-folding as a viable exomoon approach,
except for a few rare cases.

We devised an approach to solve this problem, inspired by the work of
\citet{sheets:2014}. In that work, the authors not only phase-folded each
planetary light curve but also co-added different planets together. This
allowed them to greatly increase the number of ensemble phase folded signals,
which in their case was used to boost the overall SNR. In many ways, this
approach is reminiscent of a hierarchical Bayesian model (HBM;
\citealt{hogg:2010}) but by stacking the objects the identities of each object
are sacrificed. While an HBM approach would be better suited, in general,
direct planet stacking is attractive for its simplicity, particularly if the
objective is purely to test whether an ensemble signal even exists rather than
attempting to perform detailed characterization of said signal. Therefore, in
the same vein, we decided to try stacking different phase-folded planet signals
together, to solve the $N\gg1$ problem. We highlight that \citet{hippke:2015}
independently arrived at the same idea and published before our effort, although
many differences exist in our actual implementations, as will become clear
throughout this paper.


Co-adding different planets decreases the overall noise, since we are
extending the data volume upon which our inferences are conditioned.
However, this approach is not guaranteed to increase the SNR, since many of the
objects co-added may not even possess moons and thus their inclusion only
dilutes the overall signal, rather than co-adding to it. Nevertheless, we can
quantify the overall signal amplitude as being a combination of the occurrence
rate and the moon radii. Even so, selecting a sample of planets which are
expected, \textit{a priori}, to be plausible hosts for large moons will be crucial for
maximizing our chances of a successful detection and correspondingly deriving
meaningful, physically-constraining upper limits.

\section{TARGET SELECTION}
\label{sec:targets}

\subsection{Automated Target Selection}

Not all exoplanets are equally likely, \textit{a priori}, to yield an exomoon
detection. At the most basic level, two questions guide our target selection
process: 1) what is the largest stable moon plausible around a given planet 2) 
would this moon be detectable, given the current data in hand?

In this work, we estimate whether a detectable moon is plausible following a
similar approach to that adopted in previous HEK papers. Specifically, we
employ the \textit{Target Selection Automatic} (TSA) algorithm described in
\citet{HEK2}.

To summarize, the algorithm first estimates a mass for each Kepler Object of
Interest (KOI) using the
maximum likelihood radius reported on the 
\href{http://exoplanetarchive.ipac.caltech.edu/index.html}{NASA Exoplanet 
Archive} (\citealt{akeson:2013}) and the mass-radius relation defined in 
\citet{HEK2}. This is used to further estimate the extent of the planet's
Hill sphere. Moons are expected to have their lifetime limited by the time it
takes to tidally spin out from the Roche limit to some critical fraction of the
Hill sphere, $f R_H$. Using the expressions of \citet{barnes:2002}, this logic
may be inverted to compute the maximum allowed moon mass, $M_{S,\mathrm{max}}$,
which can survive for a fiducial age of $t_{\star}=5$\,Gyr. For TSA, we set
$f=0.9309$ for the optimistic case of a retrograde moon \citep{domingos:2006}. 
Note that we assume a single moon for stability estimates and tidal evolution 
timescales; the presence of multiple moons is expected to modify these values. However,
since we do not know \textit{a priori} how many moons may be present in a given system,
it is impossible and impractical to apply more sophisticated stability estimates at this stage.

\begin{figure}
\begin{center}
\includegraphics[width=8.4cm,angle=0,clip=true]{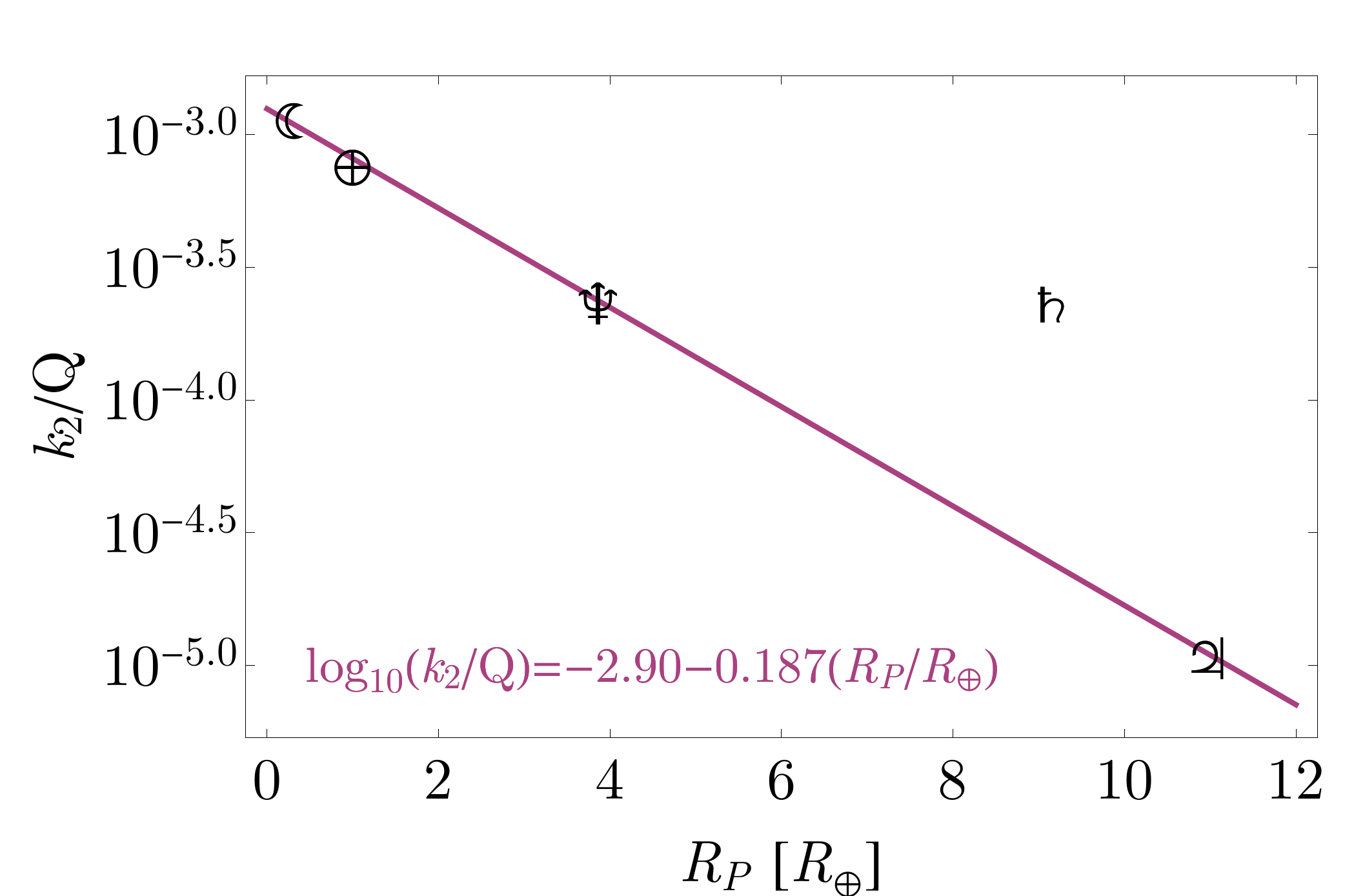}
\caption{
Relationship between the tidal property $k_{2}/Q$ and planet size for the Solar
System bodies. When excluding Saturn, the remaining four points closely follow
a power-law used in this work.
}
\label{fig:Q}
\end{center}
\end{figure}

In order to compute $M_{S,\mathrm{max}}$, we must adopt a value for the ratio 
$(k_{2}/Q)$, which represents the efficiency of tidal dissipation. Whereas in previous 
HEK papers we simply adopted $(k_{2}/Q)=0.5/10^5$, here we use an empirical 
relation based on the Solar System. As shown in Figure~\ref{fig:Q}, empirical
estimates of $(k_{2}/Q)$ for the Moon \citep{dickey:1994}, the Earth 
\citep{kozai:1968}, Neptune \citep{trafton:1974}, Jupiter \citep{lainey:2009}
and Saturn \citep{lainey:2012} follow a power-law distribution versus 
planetary radii, except for Saturn. In this work, we invert this empirical 
relation (ignoring Saturn) to estimate $(k_{2}/Q)$ for a given $R_P$. While
we don't claim this to be a fundamentally general law, it at least provides
a somewhat more reasonable estimate than the blanket fixed value assumed
previously.

Once the maximum moon mass has been computed, it is then converted into a moon
radius using the same mass-radius relation as before. Since target selection was
conducted early on in the two-year duration of this research comprising this
paper, it predated the more robust probabilistic mass-radius relation of 
\citet{chen:2017}, which is why that latter relation was not used for
these calculations. We query the combined differential photometric precision
(CDPP) of each host star \citep{christiansen:2012}, which along with the 
maximum moon radius, allows us to estimate the signal-to-noise ratio (SNR)
expected due to the moon.

For the SNR calculation, we estimate the phase-averaged signal amplitude using
the so-called ``orbital sampling effect'' (OSE) described in 
\citet{heller:2014}. While these expressions formally assume a large number
of transits, which is rarely true, they work well as an approximation for
the signal strength marginalized over the moon's phase, which is of course
unknown to us \textit{a priori}. The expected OSE flux decrease for the out-of-transit
data is given in \citet{heller:2014}, from which we may integrate over the signal to
compute the signal strength of the out-of-transit portion to be

\begin{align}
S &= \Big(\frac{R_S}{R_{\star}}\Big)^2 \frac{ \sqrt{(a_{SP}/R_P)^2-4}-2\cos^{-1}(2R_P/a_{SP}) }{\pi ((a_{SP}/R_P)-2)}.
\end{align}

The fraction on the right-hand side varies from about a quarter to a third
for $(a_{SP}/R_P)$ in the range of 5 to 100 i.e. it is a relatively gentle
function. We therefore take the limit of large $(a_{SP}/R_P)$, giving

\begin{align}
S &= \frac{1}{\pi} \Big(\frac{R_S}{R_{\star}}\Big)^2.
\end{align}

The SNR may now be calculated by dividing this by the noise expected

\begin{align}
\mathrm{SNR} &= \frac{1}{\pi} \Big(\frac{R_S}{R_{\star}}\Big)^2 \frac{ \sqrt{T_{\mathrm{Hill}}/0.25} }{ \mathrm{CDPP}_6 } \sqrt{\frac{B}{P}}
\end{align}

where $B$ is the time baseline of observations, optimistically assumed to be the
full Q1-17 baseline for these calculations and CDPP$_6$ is the combined
differential photometric precision on a 6\,hour timescale. We may express the
Hill time, assuming a simple circular orbit approximation, as

\begin{align}
T_{\mathrm{Hill}} &= f \Big(\frac{P}{2\pi}\Big) \Big(\frac{M_P}{3 M_{\star}}\Big)^{1/3},
\end{align}

which when substituted in leads to the $P$ terms cancelling out, such that

\begin{align}
\mathrm{SNR} &= \frac{ (R_S/R_P)^2 }{\pi} \Bigg[
\frac{\sqrt{2 B f} }{ 3^{1/6} \sqrt{\pi} \mathrm{CDPP}_6 }  \Big( \frac{M_P}{3M_{\star}} \Big)^{1/6} \Bigg].
\end{align}

As discussed later in Section~\ref{sec:model}, we find that the OSE model
overestimates the signal-to-noise for large $a_{SP}$, with numerical
experiments showing it is around a factor of three too high by the time we
hit $(a_{SP}/R_P)=100$. We therefore correct the SNR quoted above
by dividing by a factor of $\sim3$, which together with by the $\pi$
denominator we simply approximate to a factor of $\sim10$ denominator,
yielding

\begin{align}
\mathrm{SNR} &\simeq \frac{ (R_S/R_P)^2 }{10} \Bigg[
\frac{\sqrt{2 B f} }{ 3^{1/6} \sqrt{\pi} \mathrm{CDPP}_6 }  \Big( \frac{M_P}{3M_{\star}} \Big)^{1/6} \Bigg].
\label{eqn:SNR}
\end{align}

\subsection{Applying to the Kepler Planetary Candidates}

TSA was first run for this project in November 2014, at which time 7305 KOIs
were listed on \NEA. However, 27 were removed due to having some incomplete
column entries. Of these, 4109 were not classified as a
``false-positive'' by \NEA\ and thus were considered further. In order to 
calculate SNR, basic stellar properties are required and so we cross-referenced this list
with the \citet{huber:2014} catalog, in which we were unable to find a match for 11
KOIs. They were thus removed giving us a total of 4098 KOIs which were then put through the TSA algorithm.

\begin{figure*}
\begin{center}
\includegraphics[width=17.0cm,angle=0,clip=true]{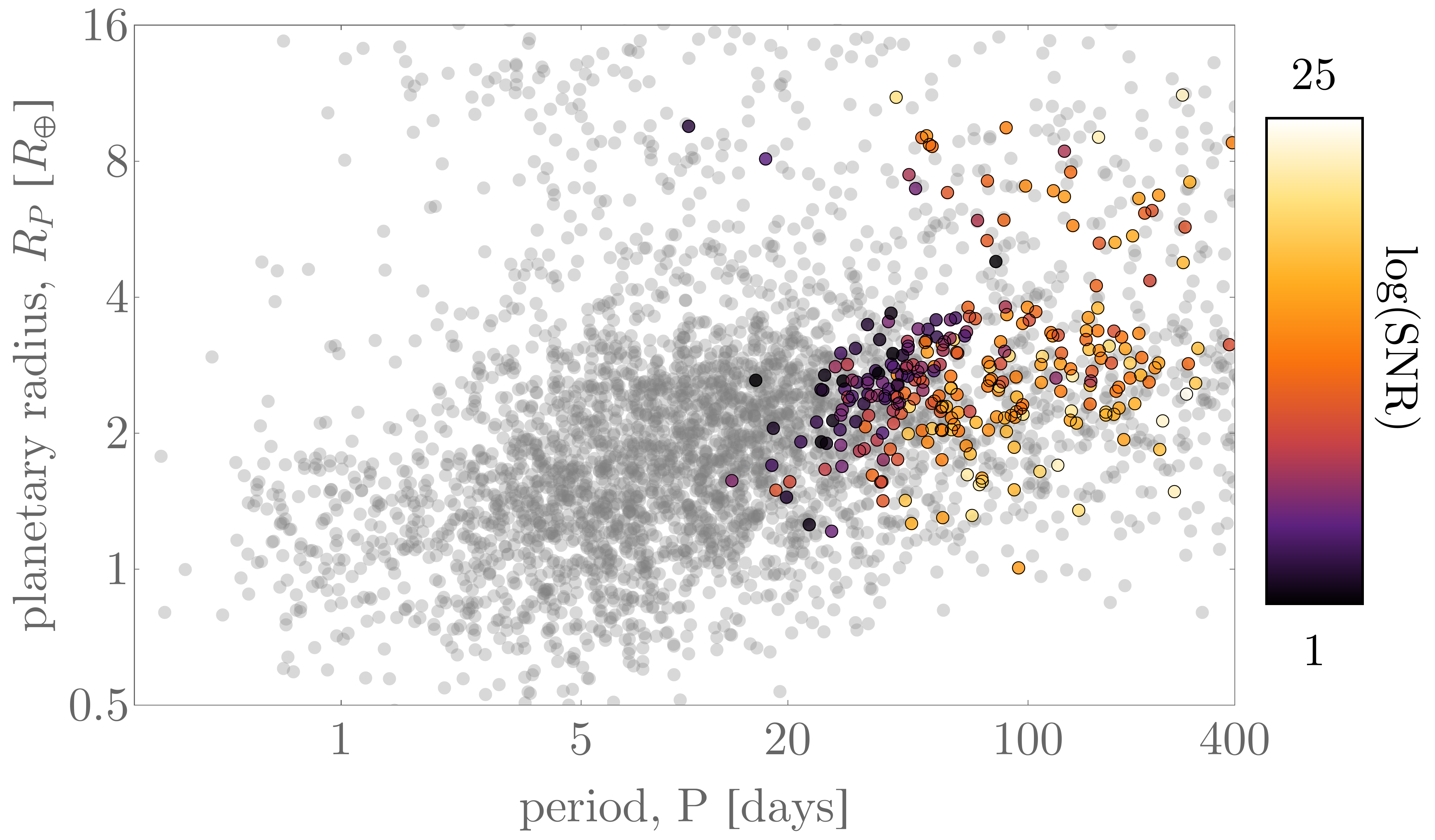}
\caption{
Location of the KOIs selected to search for exomoons around. The 966 colored
points, color-coded by the SNR given in Equation~\ref{eqn:SNR}, tend to be at
relatively long orbital periods, where Hill spheres are larger. Points with
a solid, black circle around it are the 347 KOIs found to pass our data quality
vetting.
}
\label{fig:TSAs}
\end{center}
\end{figure*}

Due to the ensemble nature of our analysis, the total SNR is expected to be
much greater than that of individual objects and thus we can afford to use
a relatively generous SNR cut. Accordingly, we elected to use SNR$>0.1$
and apply the criteria that $P<B/4=397.39$\,days (to give three
transits), leading to a sample of 966 KOIs outputted from the TSA
procedure. These targets are
visualized in Figure~\ref{fig:TSAs}.

Since this work has taken several years to complete, some of the objects
that were considered exoplanet candidates when we began are now considered false-positives.
In December 2015, we elected to remove all KOIs if they were either classified as
``false-positives'' at 
\href{http://exoplanetarchive.ipac.caltech.edu/cgi-bin/TblView/nph-tblView?app=ExoTbls&config=cumulative}{NEA}
or the probability of any false-positive scenario was in excess of 1\%, as
given by the values listed on
\href{http://exoplanetarchive.ipac.caltech.edu/cgi-bin/TblView/nph-tblView?app=ExoTbls&config=koifpp}{NEA}.
This filter removed 292 ($\sim30$\%) of the objects originally considered,
reducing the number of usable KOIs from 966 to 674.

\section{DATA PROCESSING REQUIREMENTS}
\label{sec:data_req}

\subsection{Overview}

The objective of this work is to create a phase-folded planet-stacked
out-of-transit light curve, which may be used to search
for evidence of exomoons. For the sake of brevity, we will refer to this
light curve as the \grand, or simply GLC, in what follows.

We identify four unique and critical requirements for realizing this objective,
specifically:

\begin{enumerate}
\item removal of TTVs,
\item very high quality light curves,
\item temporal rescaling,
\item two-pass data processing.
\end{enumerate}

We explain and discuss these requirements in what follows.

\subsection{TTVs}

In order to create an accurate phase-folded light curve of a sequence of
planetary transits, it is necessary to ensure the transits accurately phase up.
In the absence of transit timing variations (TTVs), this is straightforward
and is a simple linear ephemeris fold. However, the signal we seek, an exomoon,
will always introduce a small TTV signal into the data \citep{sartoretti:1999}.
Moreover, TTVs can be caused by other effects, notably perturbing planets
\citep{holman:2005,agol:2005} and thus TTVs are observed to be fairly common
($\gtrsim$10\%) in \kepler\ planetary systems \citep{holczer:2016}. Carefully
removing these TTVs is crucial in creating an accurate phase-folded transit
signal.

One approach might be to take the catalog of known TTVs from
\citet{holczer:2016} and use these for corrections. There are several reasons
why this is unsatisfactory for this work though. First, in order to assess
robust confidence limits, we require covariant, joint posteriors distributions
of the transit times and basic planet parameters, which were not derived in
\citet{holczer:2016}. Second, whenever possible, accurate phase stacking is
aided by first conducting model selection between the TTV and linear ephemeris
models, which itself formally requires computation of the Bayesian evidence -
again something not derived in \citet{holczer:2016}. Third, the TTVs derived
in \citet{holczer:2016} were conditioned on a different data set to that used
in this work. More specifically, although both \citet{holczer:2016} and this
work are based on \kepler\ photometry, our data detrending methods are distinct
meaning that these differences should be expected to affect the TTV measurements
to some degree. When one is ultimately seeking the discovery of a few
parts-per-million signal, these conventionally minor issues cannot be ignored
and should be expected to influence the results.

For these reasons, we concluded that creating an accurate \grand\ was not
possible without first deriving TTV posteriors ourselves for every system
considered.

\subsection{Data Quality}

There is a unique property of the phase-folded moon signal that has strong
implications for the data quality requirements, which is not conventionally an
issue for planet analyses. The GLC signal is a phase-fold of the planetary
transit, after removing TTVs, and thus at any given instant in phase, the moon
actually only induces a transit-dip for some fraction, $F$, of the co-added
light curves. Geometry demands that this fraction must always be less than
one-half (i.e. $F<0.5$) for all phase points occurring outside of the planetary
transit signal. This is a key point which has a major implication: median
binning kills the GLC signal.

This is extremely important, because median binning is a robust point estimate.
The forgiving nature of median binning means that one can actually do a bad
job of detrending some small fraction of your light curves (which represent
outliers) yet still recover an accurate phase-folded signal. However, if one
cannot use median binning, then one is forced to use mean-based estimates which
are sensitive to each and every transit co-added. In this case, even a single
inaccurately detrended transit light curve will contribute to the phase-stacked
signal. Once again, since we seek the detection of a signal with an amplitude
of a few parts per million, this cannot be ignored and demands the highest
levels of scrutiny and data quality.

We therefore establish that each and every transit used in our \grand\ must
be verified to be of very high data quality, which of course greatly increases
the time demands needed to complete such an analysis.

\subsection{Temporal Rescaling}

When we finally arrive at an accurate phase-folded light curve for each planet,
they must be combined into a single \grand. This is similar to the co-addition
performed for occultations by \citet{sheets:2014}. In their case, each occultation
has a distinct duration and thus simply co-adding the occultations would cause
the signal to smear out and produce an averaged signal distorted from the true
morphology. To overcome this, \citet{sheets:2014} re-scaled each event by the
known duration and then co-added, producing a more coherent signal. Just as with
the occultations, each GLC signal will have a different velocity and impact
parameter and thus cause a different duration. However, the problem is actually
worse since unlike \citet{sheets:2014}, we don't know what the true duration of
each event should be, since the duration is highly sensitive to the semi-major
axis of the moon(s), which are of course not yet discovered.

Ultimately, re-scaling will always be flawed since we can't know the semi-major
axis of the moon prior to discovering it. A full hierarchical Bayesian model
(HBM) would be an appealing way of approaching this problem, allowing each
object to have a unique semi-major axis. However, since each planet would not
satisfy $N\gg1$ transits, the OSE approximation would break down and thus each
system would require modeling with a full photodynamic simulation, such as
that from \luna\ \citep{luna:2011}. For five years, we in the HEK project have
been conducting Bayesian regression of individual systems with \luna, and the
computational demands for even individual systems are formidable
($\sim30,000$\,CPU hours per planet). Linking this into a full HBM would be
computationally extremely challenging and was not a strategy we elected to
pursue here.

Moreover, in this work, we ultimately hoped to find a signal which
was visually evident in the final \grand\ and thus not conditional upon the
inferences of an HBM. While this does not maximize the information content
of the final data product, we are motivated to follow this philosophy on the
basis that the discovery of any novel phenomenon, which exomoons would represent,
requires a much higher confidence than routine discoveries \citep{gould:2004}.

Accordingly, we seek a method of re-scaling which is ``least-bad'' and will
maximize the expected signal coherence even when marginalizing out our
ignorance of the moon's semi-major axis. One approach is that of
\citet{hippke:2015}, who re-scaled by the duration of the planetary transit.
The advantage of this is mostly simplicity, the duration is well-constrained
and easy to understand. One downside of this is that even if all the moons had
the same semi-major axes, they would still lead to the \grand\ having a smeared out
OSE signal, since each system has different barycentric velocity and impact 
parameter across the star. Another approach, re-scaling by the Hill radii, 
is not possible since the exoplanet masses are unknown. 

Instead, in this work, we argue a better approach is to re-scale the
time axis into distance from the planet, in units of planetary radii. This can be
accomplished by considering the original \citet{seager:2003} equation for the
duration of a planet, under the assumption of circular orbits:

\begin{align}
T_{23}^{14} &= \frac{P}{\pi} \sin^{-1}\Bigg( \sqrt{ \frac{ (1 \pm p)^2-b^2 }{ a_R^2 - b^2 } } \Bigg).
\end{align}

If we let $(1\pm p)\to 1$ in the above, we recover the transit duration as
defined when the center of the planet overlaps with the stellar limb,
$\tilde{T}$ \citep{investigations:2010}. Thus, at contact point 1 \& 4, we can
think of this instant in time as when a shell of radius $p$ centered
on the planet first starts to induce transit-dip features. By extension, we
could adapt $(1\pm p)\to (1+t' p)$ in the above, which would equal the duration
of a shell of radius $t' p$, centered on the planet, to start/end creating
transit-dip features. In this way, we can think $t' p$ as being the orbital
distance of the moon at the instant in time when the transit begins/ends.
Accordingly, $t'$ represents the planet-moon distance in units of the planetary
radius. This convenient form allows us to use the transit observables directly
to convert from time into a physically motivated dependent variable via:

\begin{align}
t' &= \Bigg[ \sqrt{ b^2 + (a_R^2 - b^2) \sin^2\Bigg( \frac{2\pi}{P} (t - \tau) \Bigg) } - 1 \Bigg]/p.
\label{eqn:tdash}
\end{align}

If all of the moons shared the same $(a_{SP}/R_P)$, this would produce a
coherent signal. In reality, we do not expect this statement to be true, but
moons do appear in the Solar System to be distributed log-uniformly with
respect to this term \citep{kane:2013}. This approach means that we could model
the resulting \grand\ assuming exomoons followed a formulaic distribution for
$(a_{SP}/R_P)$, such as a log-uniform.

In order to convert from $t \to t'$, we need estimates for the impact parameter
and scaled semi-major axis. Since our data is not strictly the same as that
used for the inferences quoted elsewhere, a self-consistent analysis demands
we derive these estimates ourselves, which forms another requirement for our
work.

While our conversion equation assumes a circular orbit, if we fit the data
under the same assumption, the relative estimate is at least self-consistent.
Further, eccentric planets have smaller regions of stability for exomoons
\citep{domingos:2006} and have likely experienced scattering which decreases
the chances for moons further \citep{gong:2013}. Thus, if the planet is 
eccentric, the incorrect conversion is likely irrelevant since such planets
likely do not contribute OSE-signals into the \grand\ anyway.

\subsection{Two-Pass Detrending}

In this work, we use the \cofiam\ algorithm to detrend the \kepler\
light curves. We direct the reader to \citet{HEK2} for details on the
algorithm, but essentially its goal is to remove long-term trends
without introducing any power, in a Fourier sense, at periodicities
less than the transit duration. This requirement ensures that any
signals with a time-scale approximately equal to or less than this
duration will not be distorted by the detrending process itself,
since a transit can be thought of as a Fourier series with the lowest
frequency being that of the duration \citep{waldmann:2012}. Accordingly,
both the planet and moon transits are preserved, in contrast to
polynomial-based methods which introduce power at all frequencies.

High frequency noise is not even attempted to be removed, but is monitored
by measuring the autocorrelation at the cadence-lag and used to optimize both
the harmonic filtering and subsequent identification of ``bad transits'',
which are typically rejected.

A disadvantage of \cofiam\ is the requirement for a precise estimate of
the time and duration of all transits in the time series. For this
reason, it generally is not useful for blind searches for exoplanet transits.
However, when seeking exomoons this requirement is generally true and
indeed \cofiam\ should be really thought of as an exomoon optimized
detrending method more than anything else.

As before, transit times and durations are often available for the planets
under consideration but those times and durations were conditioned on a
different data detrending. In order to make our analysis self-consistent,
we must derive these times and durations ourselves. Here-in lies a
chicken-and-egg problem though, since to derive these times we first need
detrended data, which itself first requires the times. To tackle this,
we use two passes to iterate onto the best solution. The first pass
uses the literature values for the times and duration of the transits
and then performs \cofiam. The second pass takes the times and durations
inferred using the first-pass data product, and then performs a new round
of detrending with \cofiam. This approach ensures both self-consistency
and reliability in our estimates, and provides several opportunities
to vet the data quality ensuring only the highest quality light curves
are used in the final analysis.

\section{DATA PROCESSING PIPELINE}
\label{sec:data_proc}

\subsection{Overview}

\begin{figure*}
\begin{center}
\includegraphics[width=17.0cm,angle=0,clip=true]{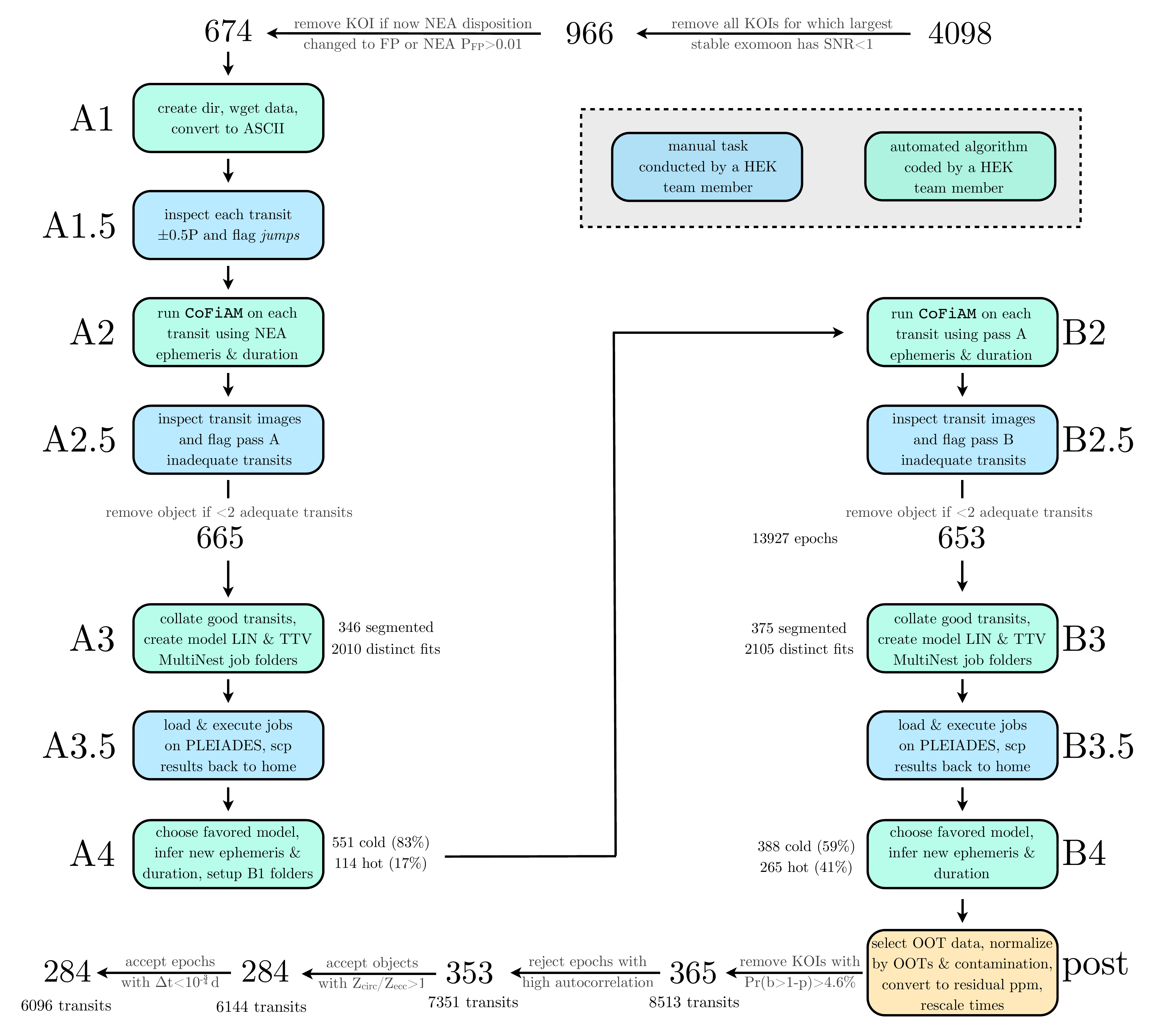}
\caption{
Schematic diagram of the pipeline used to process the \textit{Kepler}
SAP photometry to ultimately construct the \grand. We color code the manual
steps in blue and the automated steps in green. The left column broadly
describes pass A, and the right column pass B. For details on each step,
we direct the reader to the relevant subsection in Section~\ref{sec:data_proc}.
}
\label{fig:stages}
\end{center}
\end{figure*}

As motivated in Section~\ref{sec:data_req}, we require a joint posterior
distribution for the times and basic transit parameters of all planets
used for the final analysis. Using a two-pass approach to detrending-fitting
ensures that our inferences are self-consistent and conditioned upon
the actual data product used in this work. In this section, we provide a
detailed explanation of the data processing steps comprising each pass,
which are tailored to the specific and unique goals of the two. An illustrative
overview is provided in Figure~\ref{fig:stages} for reference.

\subsection{Pass A}

\subsubsection{PASS A GOALS}

We first re-assert that the primary goals of pass A are to: 1) derive transit
times for each transit 2) derive the transit duration for each planetary
candidate. These products may then be used to conduct a second detrending later
in pass B, since our detrending procedure (\cofiam) requires the times and
durations for execution.

These two objectives necessitate detrending of the data first, since we do
a covariant detrending plus inference model which require an HBM, which is
beyond the scope of this work for reasons described earlier.

For pass A, we make all decisions regarding data quality based upon whether we
conclude that these two goals can be achieved. For example, in cases where
there is plenty of out-of-transit data, but no in-transit, these would be
rejected in pass A (but can be picked up later in pass B since such signals may
still contain exomoon transits). Ultimately, defining a clear and independent
objective for each pass allows us to optimize required steps.

In what follows, we describe the different stages of data analysis which are 
performed in pass A.

\subsubsection{STAGE A1}

The first step is simply to download the \kepler\ Simple Aperture Photometry (SAP) for each
target. We wrote a shell script to step through the target list (see
Section~\ref{sec:targets}), creating a local directory for each KOI. Using
a {\sc wget} script, any and all of the target's LC data is then downloaded
from MAST (from \href{http://archive.stsci.edu/pub/kepler/lightcurves/}{archive.stsci.edu/pub/kepler/lightcurves/})
and saved in the appropriate directory. As the data were downloaded in batches,
the \kepler\ Science Operations Center data processing pipeline used for these files varied from
9.0.3 (2013-04-18 creation date) to 9.2.23 (2014-11-18 creation date).

Pre-prepared template detrending scripts are copied into each target directory.
As with previous HEK papers, we use \cofiam\ to perform our detrending. We
direct the reader to \citet{HEK2} and \citet{HEK3} for details on the
algorithm. We stress here that the algorithm is designed specifically for the
exomoon problem and requires detailed initial information such as planetary
periods, transit times and durations in order to work, which ultimately again
explains why the two-pass data processing strategy is used.

We use the \kepler\ SAP time series
throughout, since this time series has less chance of having unintended
artificial signals present than the Pre-search Data Condition (PDC) time
series \citep{stumpe:2012,smith:2012}, which has been subject to data
massaging techniques already.

\subsubsection{STAGE A1.5}

We manually inspect an image of each and every transit light curve epoch, 
centred upon the time of transit minimum expected from a linear ephemeris and
including $\pm0.5$ orbital periods of data either side. We never attempt to
stitch different quarters together and instead simply reject any data which 
occurs in a different quarter to that of the transit epoch under consideration.
As before, ephemeris parameters are taken from the NASA Exoplanet Archive for
this task. 

At each epoch, we identify if any sharp jumps, exponential flux variations or 
any other anomalous light curve feature exists in the data. This process is 
performed by one of us with the perspective of whether \cofiam\ would be 
able to fit the light curve variations or not. The aim is to keep a sufficiently 
long series of data to perform a robust detrending, but clip out patterns which 
may degrade the performance of \cofiam, with an appreciation for the basis set 
which \cofiam\ employs.

We initially pursued a variety of automated metrics for this purpose,
such as standard deviation, autocorrelation and linear trends. However, we
found that a wide variety of anomalous features survived and thus deceived
these simple metrics. Rather than creating an ever larger battery of metrics,
for which still no guarantee of completeness could be assured, we instead
acknowledged that the human eye remains an unparalleled tool in quickly
identifying anomalies in time series data.

Anomalous features are flagged by saving the instant in time just before/after
the feature, depending on whether the feature occurs after/before the time of
transit minimum. This process required approximately 60\,hours of human labor
in total.

We note that provided cotrending basis vectors are derived from the study
of common trends between stars and, in an ideal world, would provide a perfect removal
of instrumental effects. However, they do not remove stellar variations, which must also
be removed to apply our method. We therefore opt to use \cofiam, not only because it
is optimized for the moon problem, but also because it accounts for both
instrumental and astrophysical trends in a single step, which reduces the chances of artificially
injecting or removing a moon signal. Fewer manipulations of the data are preferable,
and by setting a strict frequency limit to protect 
the transit Fourier decomposition signal, we ensure that our method does not overfit out
small signals of potential interest.

\subsubsection{STAGE A2}

Stage A2 involves the first detrending of the light curves. As mentioned
earlier, this is performed automatically using \cofiam\ and the list of
anomalous features to ignore (found manually in stage A1.5). While
details of \cofiam\ can be found in \citet{HEK2,HEK3}, we point out some
general options selected for the execution in this paper.

The outlier threshold was set to 3-$\sigma$ from a 20-point moving median.
Before detrending, all planetary transits are removed with an exclusion
window of $\pm0.6T_{14}$ for all events (half a duration either side plus 20\% buffer), including the object of interest.

Each transit is detrended separately using $\pm0.5P$ of the data surrounding
each event. If the transit epoch has any associated anomalous flux changes,
as discussed in stage A1.5, then points beyond these times are also cropped. 

Periodic functions described by a sum of harmonic cosines are explored
from twice the data baseline down to twice the transit duration, with a
cap of 30 harmonics (beyond which we tend to encounter numerical
instabilities). This choice ensures that \cofiam\ does not disturb the shape
of the planetary transit in a Fourier sense \citep{waldmann:2012}.

Each model is regressed to the data then ranked by the local autocorrelation,
as computed using the \citet{DW:1950} statistic (DW) on the timescale of the LC
cadence, with the lowest autocorrelation being favored. The favoured model is
then applied to original time series, re-including the planetary transit of 
interest. The final light curve is saved with a $\pm6.66T_{14}$ gap either
side (which is the local timescale used for the DW calculation). 

Finally, an image of every detrended transit is stored along with the best DW
statistic.

\subsubsection{STAGE A2.5}

This is the second manual stage in pass A, where we manually identify ``bad''
transits. In some rare cases, \cofiam\ fails and produces a light curve which
cannot be used for fitting, due to visually evident trends remaining in the data.
For example, if we missed a location of an anomalous flux variation in stage
A1.5, \cofiam\ may be trying to detrend sharp features with a smooth cosine
function, producing a poorly detrended light curve. By manually going through
the light curves in this way, it is essentially a second-check of the data
quality, catching any missed anomalies from earlier. 

In general, these assessments are made by one of us searching for any 
visually evident trends which would significantly impede our ability to fit the
light curves to determine $T_{14}$. A bad transit does not necessarily have a 
poor DW statistic, although that tends to be a common scenario. Because we 
anticipate a second  pass, we can be generous in considering acceptable data 
qualities at this stage. This acceptance level is non-constant, since we try to 
allow KOIs displaying very frequent bad transits to have at least a few transits 
which can be used for fitting in stage A3. Vice versa, if a KOI has many clean
transits, we apply more stringent conditions in assessing data quality. Finally,
we note that assessments are generally based on SNR of the transit, not the
raw wiggles in the data, but the relative size of those wiggles compared to the
transit. In cases of very low SNR, where the transit is not visible in a single
epoch, we work under the assumption that sometimes data will wiggle up and
sometimes down, but we must trust that on the average there is power (thus
we try to allow for almost anti-transit like features in the interests of being
unbiased and balanced). In such cases, our criteria switches from trying to make
a good measure of the transit duration to simply avoiding ``catastrophic''
detrendings.

In addition to a ``bad-transit flag'', we also use a ``sparse flag'' for
transits where there is insufficient in-transit data (or none at all). In
some rare cases, a third type of flag was used, ``missing flag'', where the
data are well-detrended, we have good temporal coverage, but a transit which
should be visually obvious (given the transit depth) is missing in the data.
We consider these cases to be most likely due to an erroneous transit
ephemeris on the \NEA.

If fewer than two good transits remain for a KOI, the object is removed
from our sample as being a useful object. In total, this removes 9 KOIs
dropping our sample down from 674 to 665 KOIs. It should be noted
that any rejection of bad data, manual or automated, injects additional 
uncertainty into the occurrence rate of exomoons calculated in this work 
which is, strictly speaking, of unquantifiable magnitude. If there is any 
correlation between unusable transits and the presence of moons the calculation
could be particularly affected, but this is impossible to measure since the bad transits are
by definition unusable for exomoon characterization. We proceed under the 
assumption that bad data result from instrumental effects that are distributed randomly across
the data set.

\subsubsection{STAGE A3}

The third-stage is an automated shell script which begins by stitching the good
transits together for each KOI into a single file. Our script then creates two
directories for a linear ephemeris model fit, $\mathcal{H}_{\mathrm{LIN}}$,
and a transit timing variation model fit, $\mathcal{H}_{\mathrm{TTV}}$, to be
fed into \multi\ \citep{feroz:2008,feroz:2009}. The script then queries the
orbital period, $P$, and time of transit minimum, $\tau$, from \NEA\
database to construct priors for these terms. In the case of model
$\mathcal{H}_{\mathrm{LIN}}$, the prior on both terms is uniform centered on
reported \NEA\ value with $\pm1.0$\,days window. For model
$\mathcal{H}_{\mathrm{TTV}}$, the period is treated as a fixed parameter with
the individual transit times following a uniform prior centered on the
expected time of transit for a linear ephemeris with a $\pm1.0$\,day window
again. The choice of the 1.0 day window is essentially arbitrary, but assumes that
transit timing variations larger than one day are highly unlikely in the region 
between 0.1 and 1.0 AU (c.f. \citet{holczer:2016} who find $< 0.3\%$ of their sample
have TTV amplitudes larger than one day). A larger window will unnecessarily increase the time it takes to explore
the parameter space, while a smaller window tailored to the linear ephemerides of each planet 
could result in a bias against finding large TTVs.

For model $\mathcal{H}_{\mathrm{LIN}}$, the basic free parameters are
$P$ \& $\tau$, two quadratic limb darkening terms (we use the $q_1$ \& $q_2$
prescription from \citealt{LDC:2013}), the ratio-of-radii, $p$, the impact
parameter, $b$, and the stellar density, $\rho_{\star}$. Uniform priors
are adopted for all except $\rho_{\star}$ which follows a log-uniform
prior. This gives a total of $d=7$ free parameters, which is easily handled by
\multi. For model $\mathcal{H}_{\mathrm{TTV}}$, the period is treated as
fixed, removing one degree of freedom, but then each transit epoch has a
unique $\tau$ parameter (same prior as $\mathcal{H}_{\mathrm{LIN}}$),
giving us $d=6+N$ free parameters. For $d\gtrsim20$, the performance of
\multi\ is severely impeded and global fits are not possible.

A common approach is ``templating'', where one folds the transits,
creates a template which is then regressed to the individual epochs
(e.g. \citealt{holczer:2016}). This approach underestimates measurement
uncertainties since it ignores the covariance between the transit shape
parameters and the individual transit times. Rather than underestimating errors, 
we prefer to overestimate them and so adopt a different strategy. 
Instead, we split up our light curves in segments of $\sim10$ epochs
each, providing a manageable number of free parameters for each. The downside
is that each segment is not able to utilize the information about the global
transit shape learnt from other segments, and so the uncertainties will
be larger (but more robust) than templating. Accordingly, in stage A3 our
script automatically segments the data up where necessary.

We find a total of 346 out of our sample of 665 KOIs require segmenting,
whereas the rest are treated in a single fit due to the tractable number
of epochs available.

\subsubsection{STAGE A3.5}

Stage A3.5 was primarily performed using NASA's Pleiades cluster, and
essentially involved loading, compiling, executing and then retrieving the
over two thousands light curve regression jobs required. In total, we
estimate that approximately $\sim100,000$ CPU hours were used during this
phase of the analysis and spanned several months of wall time.

\subsubsection{STAGE A4}

In the fourth stage, we segue into pass B by laying the ground-work needed
using the results from our light curve fits. The first task is to decide for
each KOI whether it is dynamically ``hot'' or ``cold'', by which we mean
whether $\mathcal{H}_{\mathrm{LIN}}$ model (cold) or
$\mathcal{H}_{\mathrm{TTV}}$ model (hot) is preferred. In cases where the fits
were completed using a single segment, the evidences from \multi\ can be
directly used to compute the Bayes factor and rigorously assign the preferred
model. For segmented models, direct evidence calculation is not possible
since the TTV model has multiple copies of the same parameters for the transit
shape. Instead, we use weighted linear regression to find the maximum
likelihood linear ephemeris through the posterior transit times and then
inspect the residuals for evidence of TTVs. This is simply done using a
$p$-value test searching for a excessive $\chi^2$ (cut off used was
3\,$\sigma$).

Formally, model assessments using a $p$-value are incorrect since they are never
actually compared to another model. More precisely, the $p$-value test
is prone to inferring significant evidence for the alternative
hypothesis even in cases where it should not. For example, excess noise from
other sources or a single poor measurement could lead to the $p$-value test
favoring the TTV model erroneously. Let us consider the effect of this by
imagining a linear ephemeris fit to a set of transits with the TTV model. The
times of transit and basic transit parameters will all still come out formally
correct, just with inflated uncertainties. Giving each epoch a free transit
time is still able to recover the original linear ephemeris solution.
Therefore, despite the $p$-value's tendency to overestimate significance, this
merely acts to conservatively inflate our uncertainties and does not formally
invalidate our inferences.

In total, we find 551 of the 665 KOIs are ``cold'', with the remaining 114
being dubbed hot. For comparison to later, we point out that one might expect
pass B to increase the hot fraction due to the improved detrending and thus
greater sensitivity to even small TTVs.

Stage 4 ends by duplicating all of the KOI folders into a new directory with a
small text file recording the favored model and ephemeris parameters. The
transit duration is also recorded in this file, where for
$\mathcal{H}_{\mathrm{LIN}}$-favored KOIs is computed directly from the joint
posteriors and for $\mathcal{H}_{\mathrm{TTV}}$-favored KOIs from a weighted
sum of each segment's marginalized credible interval for the duration.

\subsection{Pass B}

\subsubsection{PASS B GOALS}

Before describing each data processing step for pass B, we first
re-assert the objectives, which represents the backdrop against which
all decisions in pass B are framed.

Ultimately, the data product from pass B should be high-quality, cleaned
light curves with accurate estimates of the transit times and basic parameters
needed for stacking and re-scaling. Our tolerance for poor-quality light
curve is necessarily more stringent here, since unlike pass A, there is
second-chance for these light curves and they have to be of sufficient
quality for stacking by the time pass B is complete.

\subsubsection{STAGE B2 \& 2.5}

Downloading the data (stage 1A) and removing jumps (stage A1.5) does not need
to be repeated since the raw data product is unchanged. Accordingly, we
skip straight to stage B2. Mirroring stage A2, we detrend the SAP light curves
using \cofiam\ but now using the duration and transit times determined earlier
in stage A4 (specifically we use the maximum \textit{a posteriori} values).

In stage 2.5 we again inspect these light curves visually for poorly detrended
examples and find 12 KOIs end up with fewer than two usable transits after this
process. Removing these objects reduces our sample from 665 to 653 KOIs.

\subsubsection{STAGE B3 \& 3.5}

As with stage A3, stage B3 collates the good transits and sets up folders for
models $\mathcal{H}_{\mathrm{LIN}}$ and $\mathcal{H}_{\mathrm{TTV}}$ ready
for fitting with \multi. Once again this results in just over two thousand
distinct jobs to run, with 375 of the KOIs being segmented.

Unlike stage A, we here allow the light curve model to account for any
known blending from nearby sources. These are collated from
\citet{everett:2015,kolbl:2015,adams:2012,adams:2013,dressing:2014,law:2014},
and we use \kepler-converted magnitudes to estimate the appropriate
contamination factor for each band. These blending factors are treated as
Gaussian priors, with a standard deviation set by the uncertainty on the
converted magnitude. In total, 39 of the targets required a blending term
to be included.

In stage 3.5, we again load, compile, execute and retrieve these runs on the NASA 
Pleiades cluster, requiring another round of 100,000 CPU hours and several months
of wall time.

\subsubsection{STAGE B4}

Finally stage B4 repeats stage A4 with the new light curve fits, performing
model selection using the same framework described earlier. After completion,
we found that the fraction of cold KOIs indeed decreased as expected from
83\% to 59\%, giving 265 hot KOIs in our sample. In this work, these TTVs
represent purely a nuisance but we acknowledge that this data set represents
a rich and interesting catalog for others in the community. We therefore
make all of the transit times, both for the hot and cold samples,
publicly available at \url{github.com/alexteachey/TTV_posteriors}.

For each KOI, we export the maximum \textit{a posteriori} transit fit and the
corresponding vector of out-of-transit baseline fluxes (OOTs), which are found
by linear minimization of the maximum \textit{a posteriori} model and the data (this is
actually done on the fly during every regression step, following the approach
using by \citealt{kundurthy:2011}). These OOTs will be useful later for
stacking the final light curves.

\subsection{Post-Processing}

In post-processing, we aim to export a single file for each KOI which contains
a phase-folded light curve suitable for planet-stacking. First, the transit
times are removed using the favored model and the maximum \textit{a posteriori} 
parameter vector. Next, the planetary transit is removed by excluding
all data which falls within the 2\,$\sigma$ upper limit of the derived full
duration, $T_{14}$. Each epoch is then corrected for any residual DC power
detected by the OOT vector regressed during Stage B4. Global blending factors,
as well as quarter-to-quarter aperture flux contamination factors are corrected
for following the approach described in \citet{nightside:2010}.

Next, we subtract unity from the normalized fluxes and multiply by a million
to create a ppm residual light curve for each object. Finally, the time column
is converted to $t'$ using Equation~(\ref{eqn:tdash}) and the maximum
\textit{a posteriori} transit parameters derived from the preferred model regressed
in Stage B4.

\subsection{Filtering}

Before stacking these planets together, we first remove KOIs and individual
transits which fail to satisfy several criteria. First, we require a
2\,$\sigma$ confidence that the planet is not a grazing event, which would
mean that $b>(1-p)$. Grazing events have degenerate planetary radii and
could be far larger in size, potentially even a false-positive. Erring on
the side of caution, we remove any such KOI which filters out 288 KOIs,
leaving us with 365 targets.

We next test for excessive autocorrelation using the DW metric.
For each transit, we generate 10,000 mock realizations where the
data are drawn from perfect normal distributions at the exact same
sampling observed in the data and the reported uncertainties. We
use these synthetic transits to generate an expected distribution for
the DW metric and flag any transits for which the real DW metric is
more than two standard deviations away from our synethetic population.
If a KOI has 50\% or more of it's transits flagged in this way,
the entire KOI is dropped from the sample. This filtering removes
a further 12 KOIs.

Next, we compare the transit derived stellar density (which assumes
a circular orbit) to an independent estimate, in order to exploit the
photoeccentric effect \citep{MAP:2012,dawson:2012} to infer a
minimum eccentricity, $e_{\mathrm{min}}$, of each KOI \citep{AP:2014}.
We draw a random sample from the transit derived posterior found in stage
B4 and divide it by a random sample drawn from the KOI's corresponding
stellar density posterior derived in \citet{mathur:2017}. This ratio is
then converted into a minimum eccentricity using Equation~(39) of 
\citet{AP:2014}, and the process is repeated until we have derived
40,000 posterior samples for $e_{\mathrm{min}}$ for each KOI. For
each KOI, we also construct a prior for $e_{\mathrm{min}}$ based off
the prior used in the transit fits and \citet{mathur:2017} distribution.

\begin{figure*}
\begin{center}
\includegraphics[width=18.0cm,angle=0,clip=true]{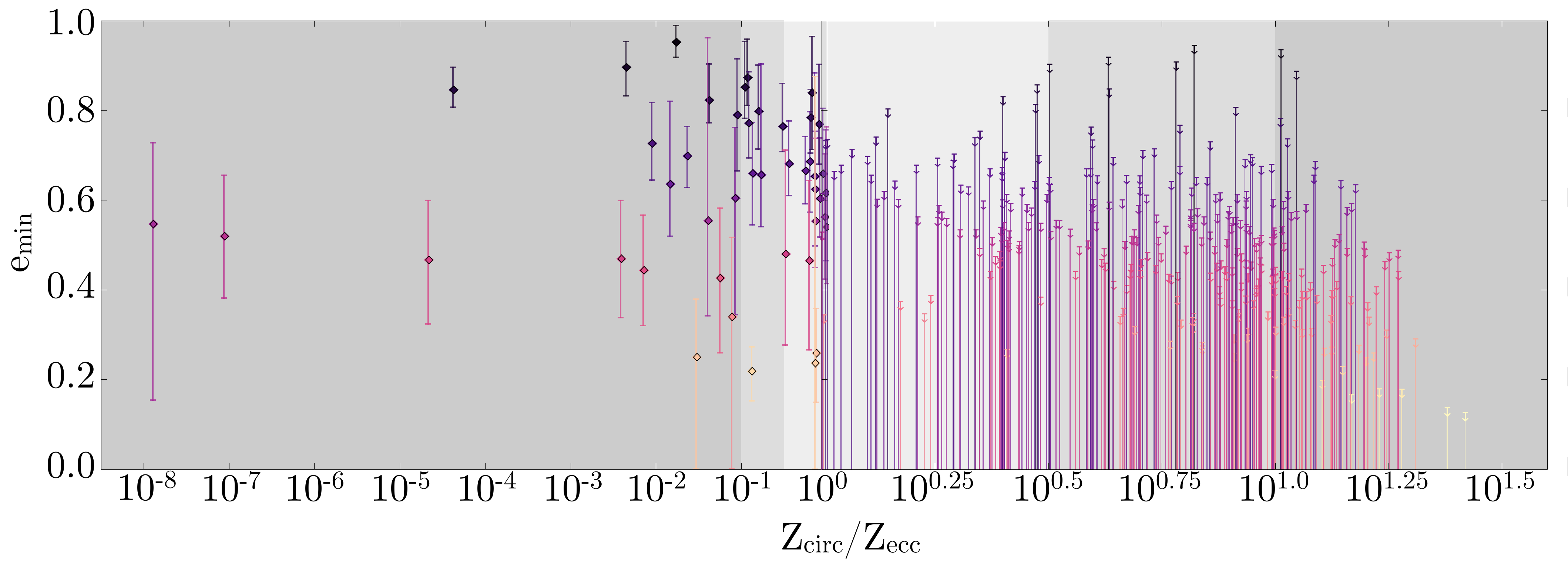}
\caption{
Minimum eccentricity of 353 KOIs derived using the photoeccentric
effect as a function of the Bayes factor for a circular vs eccentric
orbit. KOIs to right of unity are depicted as upper limits on
eccentricity, whereas we plot 1\,$\sigma$ credible intervals for the
others. The 284 KOIs favoring a circular orbit are considered further
as viable exomoon candidates in this work.
}
\label{fig:photoeccentric}
\end{center}
\end{figure*}

We next evaluate the Savage-Dickey ratio between the posterior and
the prior to estimate the Bayes factor,
$Z_{\mathrm{circ}}/Z_{\mathrm{ecc}}$. We find that 284 KOIs have
$Z_{\mathrm{circ}}/Z_{\mathrm{ecc}}>1$, implying a near-circular orbit,
whereas the other 69 KOIs are rejected for further analysis, on the
basis that eccentric planets likely result from scattering which would
disrupt moon systems \citep{gong:2013}. We plot $e_{min}$ as a function
of the Bayes factor in Figure \ref{fig:photoeccentric}.

Finally, we elected to remove transits for which we are unable to
measure the transit time to within a precision of $10^{-0.75}$\,days,
chosen to remove unconverged posteriors given the prior width,
which is necessary to ensure we are able to reasonably phase-fold
transits together. This did not change the number of KOIs from 284,
but did reduce the number of transits in our sample from 6144 to
6096.

\subsection{Constructing a \grand}

With each target now having a fully processed phase-folded light curve,
we are finally ready to stack different targets together to create a
\grand. This stacking can be across all 284 surviving targets, or
a subset of them, as explored later. Although we describe here the case 
for the complete ensemble, the planet-stacking methodology is the same 
when dealing with subsets.

Across the 284 KOIs, we have 6096 unique transits comprising of 364059
photometric measurements. The re-scaled times are well-described by a
half-normal distribution with a standard deviation of 113. We elect to
remove any points which fall outside of the range $t'>150$, leaving us
with 309750 points.

\begin{figure}
\begin{center}
\includegraphics[width=8.4cm,angle=0,clip=true]{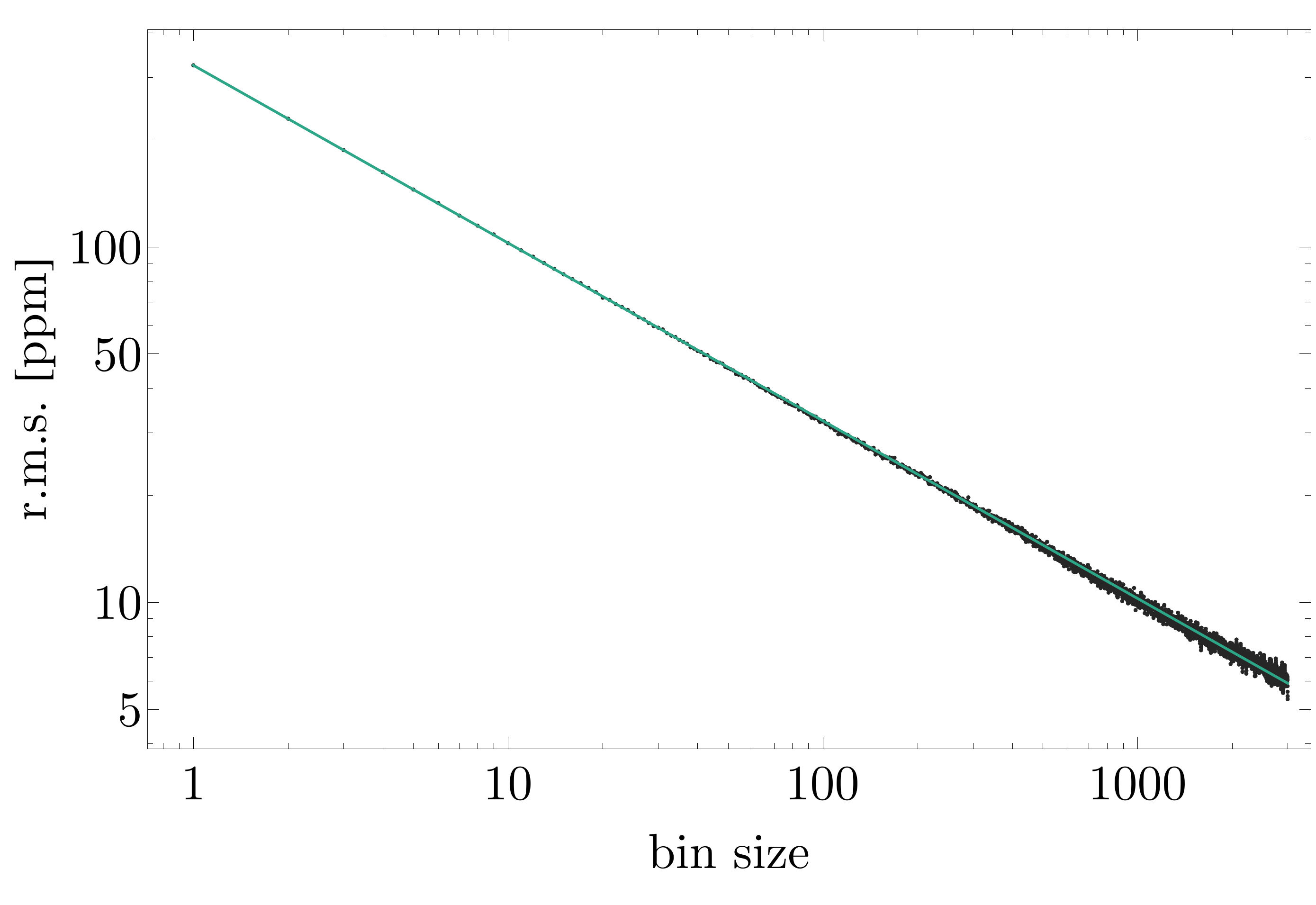}
\caption{
RMS of the \grand\ as a function of bin size, demonstrating the
expected $N^{-1/2}$ scaling of white noise.
}
\label{fig:rms}
\end{center}
\end{figure}

The \grand\ photometry shows no evidence for correlated noise structure,
as expected from averaging so many independent data sets together. This is
verified in Figure~\ref{fig:rms}, where we plot the root mean square
(r.m.s.) of the time series as a function of bin size, which displays
excellent agreement with the expected root $N$ scaling.

Dividing the fluxes by their formal reported uncertainties, we find
that the robust r.m.s. (given by 1.4286 multiplied by the median
absolute deviation) equals 1.09, indicating only a small amount of
extra noise above the reported uncertainties. We re-scaled the errors
by this factor and then performed 3\,$\sigma$ clipping, removing just
under one percent of the points. The final time series is found to have
a standard deviation of 5.1\,ppm when binned to a scale of $\Delta t'=0.5$.

\section{MODELING}
\label{sec:model}

\subsection{Choosing a modeling formalism}
\label{sub:OSEtests}

In addition to manipulating the \kepler\ data to construct the grand light
curve, we also require the ability to model its shape, as a function of
various exomoon parameters of interest.

There are two possible avenues to modeling the grand light curve. The first
is to model the individual systems then combine them to create an ensemble
model, and the second is to use a model describing the ensemble from the
outset. The latter approach describes the model proposed in \citet{heller:2014}
and later modified in \citet{heller:2016}, who refer to this model as
the ``orbital sampling effect'' (OSE). The great advantage of this approach
is that one can employ analytic expressions described the ensemble signal
without having to laboriously simulate each of the individual systems
and then combine later. Thus, in principle, the OSE approach has the advantage
of speed and being more straight-forward in application. Indeed, this was
the model used in \citet{hippke:2015}.

The alternative approach would be to use a detailed ``photodynamical'' light
curve simulator, such as \luna\ \citep{luna:2011}, to predict the light curve
of each system with some trial set of moon parameters and then later combine
them to produce a grand model. Photodynamics, a test first coined in
\citet{carter:2011}, refers to a light curve simulator which evolves a
planetary system at each time step and computes the corresponding shadows cast
onto the sky-projected stellar disk. Unlike the OSE model of
\citet{heller:2014}, this model is not specific to phase-folded events but of
course can be easily used to simulate such a case by simply folding the final
predicted light curve.

In general, \luna\ provides a physically detailed light curve simulation, but
comes at the expense of greater computational cost than the simple closed-form
expressions of OSE. For these reasons, if the accuracy of OSE is validated,
it would be far simpler and thus preferable to employ the OSE formalism for
our model fits of the grand light curve. However, after photodynamical testing
of the OSE predictions and consideration of the specifics of our problem hand,
we came to the conclusion that OSE would not be an accurate modeling tool for
our data product. In particular, we argue that three key barriers prevent us
from directly using the OSE models to describe our grand light curve:

\begin{itemize}
\item Inter-population variation,
\item Heteroscedastic weighting,
\item Laplace resonances.
\end{itemize}

We briefly describe these three reasons in what follows.

\subsubsection{Inter-population variation}

The OSE model is derived assuming one co-adds many transits of the same
planet-moon system i.e. that the basic parameters of the system are not
changing. However, in our case we co-add different systems together which
have distinct planet and moon parameters. For example, in our Galilean
moon fits, we assume that the moons have inclinations representative of
Io, Europa, Ganymede \& Callisto, but each planet's moon system will have
unique moon inclinations, despite being drawn from a common underlying
distribution. While we could co-add many OSE signals together modeled
individually with the corresponding parameters, each OSE curve would be
modeling only a small number of transits and thus would be formally
invalid - since it is by definition an ensemble model. Without detailed
investigation, it was unclear that one could simply co-add across a
population in this way and recover the correct phase stacked signal and
thus we preferred to use \luna\ which provided an accurate model of the
individual events.

\subsubsection{Laplace Resonances}

A subtle and minor point of concern was dealing with the Laplace
resonance in the OSE framework when modeling Galilean analogs. In \luna\,
each individual moon transit is generated and thus we are able to assign
relative phases between the satellites such that they reside in
not only the correct mean motion resonance but also share the Laplace
resonance in terms of their mutual phases:
$\pi = \lambda_{\mathrm{Io}} + 3\lambda_{\mathrm{Europa}} + 
2\lambda_{\mathrm{Ganymede}}$. In contrast, the OSE framework never models
the individual events, rather just the ensemble, and we were unable to
demonstrate that OSE was correctly accounting for such a phase lock.

\subsubsection{Heteroscedastic weighting}

Finally, OSE is an average of light curves, which by definition means
each light curve is given precisely the same weight. Second, each light
curve contribution is assumed to be uniformly and densely sampled. Our real
data products do not satisfy such constraints, since first we co-add
the different planets together using weights based off the root mean
square of the photometric residuals. Second, light curves are not
uniformly sampled, featuring data gaps and removed outliers, as well
as being non-uniformly transformed in time via our temporal re-scaling.
Since \luna\ models individual events, data gaps, integration time
effects \citep{binning:2010} and re-scaling are easily accounted for before
applying any co-addition, enabling us to ensure our model is representative
of the data.

\subsection{Photodynamic Look-Up Tables (LUTs)}
\label{sub:LUTs}

As a result of the myriad of complicated effects influencing the final
model yet the relative low-dimensionality of the model itself, we elected
to build a grid of pre-computed models which accounted for all of the
effects described above. For each KOI, we took the maximum \textit{a posteriori}
transit parameters from the planet-only fits (using the favored model)
and generated a planet+moon light curve using \luna\ with the same planet
parameters but adding in one or more moons. The model curve is generated
at precisely the same cadence as the data used for the planet-stack
and accounts for the long-cadence integration time using 1 minute numerical
re-sampling \citep{binning:2010}. After all of the KOI model light curves have
been computed for a specific choice of underlying moon population, they
are co-added with the same weighting used for the real data. In other words,
we inflict precisely the same transformations to the model as we do to the
data, to ensure a like-for-like comparison at the end.

\subsubsection{Galilean Analogs}

The moons are generated in two ways. The first case was for a Galilean analog.
Here, we assume that four moons orbit each KOI
with properties resembling those of Io, Europa, Ganymede \& Callisto. To
inject some stochastic variation between each moon, yet maintain the
1:2:4 resonance of the inner three, we randomly place each Io-analog to
have a semi-major axis of $(a_{SP}/R_P)\sim\mathcal{U}[0.8\times6.1,1.2\times6.1]$,
where $6.1$ is the actual value for Io around Jupiter and
$\mathcal{U}[a,b]$ is a uniform distribution from $a \to b$. The next moon
along is then assumed to lie in a 2:1 resonance, such that
$(a_{SP}/R_P)_{\mathrm{Europa}} = 2^{2/3} (a_{SP}/R_P)_{\mathrm{Io}}$, and
similarly for the Ganymede-analog with respect to the Europa-analog. The
semi-major axis of the Callisto-analog, which does not reside in the resonance
chain, is then randomly drawn as $(a_{SP}/R_P)_{\mathrm{Callisto}} \sim
\mathcal{U}[3+a_{SP}/R_P)_{\mathrm{Ganymede}},1.2\times27.2]$, where $27.2$
is the actual value for Callisto around Jupiter. Any moon systems generated where
two moons have semi-major axes within 3\,planetary radii of each other are
rejected.

\begin{figure}[!]
\begin{center}
\includegraphics[width=8.4cm,angle=0,clip=true]{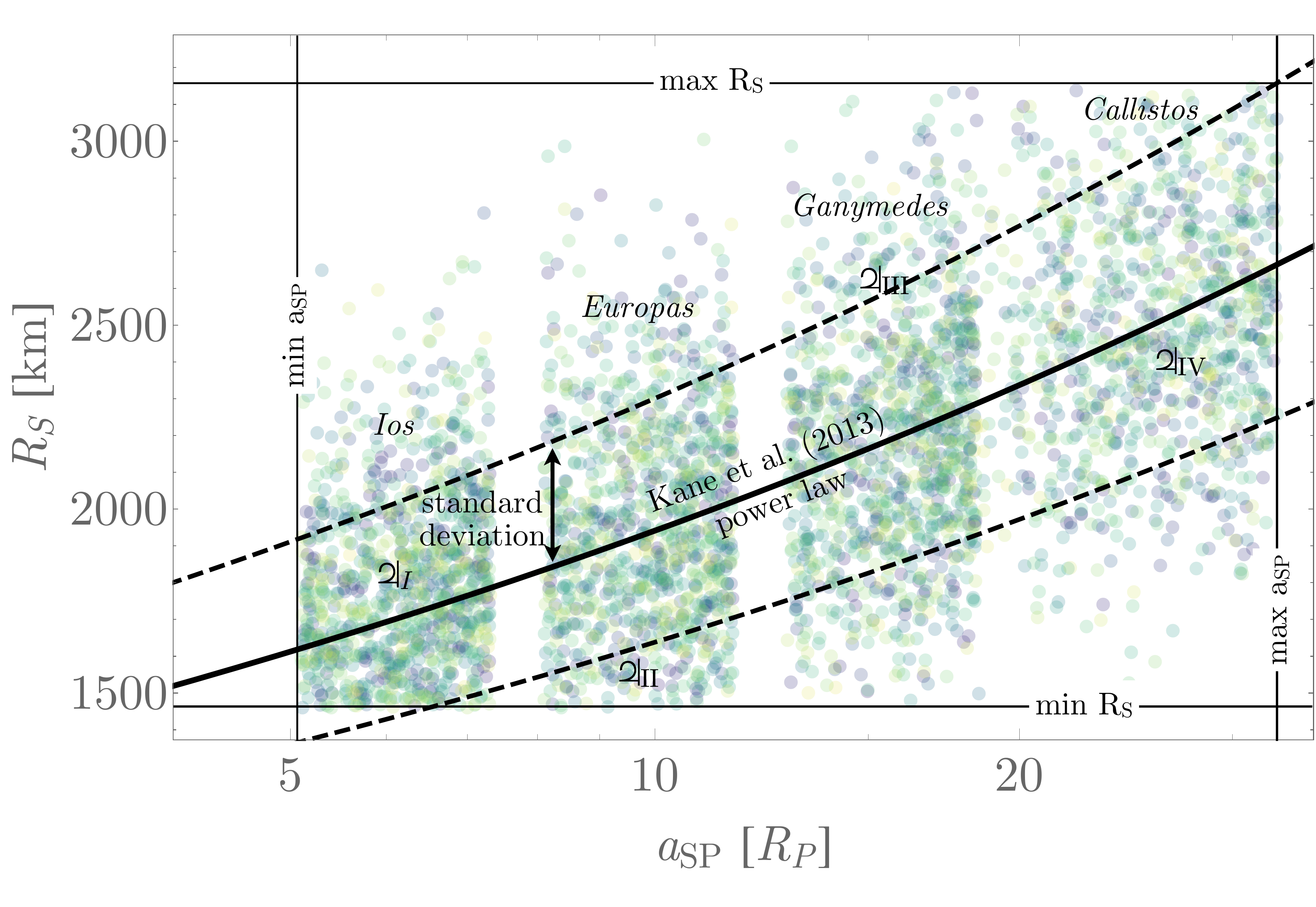}
\caption{
One thousand randomly generated Galilean-moon system analogs using the
method described in the main text. Each color represents a unique system
comprising of four moons.
}
\label{fig:galilean_distrib}
\end{center}
\end{figure}

To generate moon radii which were stochastic yet representative, we adopted the
radius-separation power-law model of \citet{kane:2013}. The authors note that
the radii of moons tend to increase with respect to semi-major axis following
a power-law model. We took the four Galilean moons in isolation and calibrated
a least squares power-law model to it, giving $\log(R/R_{\oplus}) \sim \mathcal{N}(6.95 +
0.27 \log(a_{SP}/R_P), 0.17)$, where the standard deviation quoted is that
resulting from the residuals of the best-fitting line. To protect against
peculiar draws, we required that the quadrature sum of the radii was within
20\% of the actual sum for the Galilean system, that the minimum radius
moon was at least 80\% the radius of Io and the maximum radius moon was
no more than 120\% the radius of Ganymede. The resulting covariant distribution
is illustrated in Figure~\ref{fig:galilean_distrib}.

To go from physical radii, $R_S$, to $R_S/R_P$ needed for the modeling (or
equivalently $R_S/R_{\star} = (R_S/R_P) (R_P/R_{\star})$), we used planetary
radii derived in \citet{chen:2017b} to make the conversion. We also included
the slight TTV and TDV effects induced by the Galilean moons by including
$(M_S/M_P)$, which was computed by using {\tt forecaster} \citep{chen:2017}
applied to the moon radius to predict a mass, and then using the physical
planetary masses predicted in \citet{chen:2017b}.

The mean longitude of the inner two moons are randomly generated uniformly but
the third is enforced to satisfy a Laplacian resonance. The inclination of the
moons are randomly drawn from a Von Mises distribution with $\kappa=42637$,
which we found to maximize the likelihood of a Von Mises distribution
conditioned on the real Galilean moons. Eccentricities were kept fixed at zero
and the moons all follow perfect keplerian orbits i.e. we do not model
gravitational interactions between the moons.

After stacking the resulting model light curves with correct cadence and weightings,
we tried varying the fraction of systems which harbor moons. Treating each system
as a Bernoulli experiment with a probability of having a moon system given by
$\eta$, we found varying $\eta$ was equivalent to simply scaling the $\eta=1$
resulting light curve by the same factor. Having demonstrated this, we were able
to exploit it to aid in later fits.

In cases where a subset of systems were modeled, the process described above was
repeated creating unique model light curves for each specific subset. Generally,
the shape of each resulting light curve were very similar, but ended up with
different amplitudes as a result of the differing weights, data gaps and stellar
radii for the host stars. As noted these factors represent data specific properties
for each subset and were saved for later use with the single moon simulations.

\subsubsection{Single Moons}
\label{subsub:single_moon_LUT}

While a single moon is four times quicker to generate than four Galilean moons,
the Galilean moons follow an expected distribution in terms of their sizes and
orbital semi-major axes. For a single hypothetical moon, we have no idea what
these properties are \textit{a priori} and thus our grid cannot simply span $\eta$, as
before, but now must also span $R_S$ and $a_{SP}$ leading to a three-dimensional
look-up table. Fortunately, the effect of $\eta$ is a simple scaling and thus
can be applied easily during the fits themselves, yet this still means we need
to generate a two-dimensional grid of models, rather than just a single look-up
example in the case of the Galilean-analogs.

We setup a logarithmically-spaced grid from $R_S=0.2$\,$R_{\oplus}$ to
$R_S=2.0$\,$R_{\oplus}$, with 16 unique grid points. For $a_{SP}$, our
grid is again logarithmic, defined as $a_{SP}=2^x$ where $x$ is stepped
through from 1 to 7 in 0.1 steps, leading to a total grid size of 976
elements. The moon is treated as being exactly coplanar and circular with
random phase and as before we generate unique light curves for each KOI
and then co-add with the appropriate weightings to create our final models.
For these simulations we set $(M_S/M_P)=0$, since some simulations
permit very massive moons which would cause noticeable TTVs, which would then
be subsequently removed anyway by our data processing pipeline described
earlier.

When dealing with subsets, we apply the scaling factors found earlier with the
Galilean-analog experiments, since the computation time to create the grid
required many weeks. During the actual fits, we used a bi-linear interpolation
of every unique binned photometric data point, conditioned upon the LUT. We
also added an extra grid point of $R_S=0.0$\,$R_{\oplus}$, corresponding to a
flat line, to provide numerical stability if fits attempted to compute the
likelihood of radii below our $R_S=0.2$\,$R_{\oplus}$ limit.

\section{ANALYSIS}
\label{sec:analysis}

\subsection{Galilean Global Fits}
\label{sub:galileanfits}

We first discuss our results from regressing our photodyanmical phase-folded
planet-stacked planet+moon light curve models (see Section~\ref{sec:model})
to all 284 KOIs deemed to be of suitable quality for our analysis (see
Section~\ref{sec:targets}). As discussed, the fits are conducted for two
different light curve models, a Galilean-analog and a single moon.

For the Galilean-analog, the only parameter directly affecting the light curve
model is $\eta$, the fraction of KOIs which harbor a Galilean-analog. In
addition, we added two other free parameters into our fits. The first was to
account for excess photometric scatter, $\sigma$, and was simply added in
quadrature to our derived uncertainties in the planet-stacked light curve. The
second was an offset term, $\gamma$, to allow for a re-normalization of the
data set. While $\eta$ and $\gamma$ were assigned uniform priors, $\sigma$
was assumed to follow a log-uniform prior from 0.1 to 10\,ppm. The regression
was performed using \multi\ with 1000 live points. 
The fits were repeated ten times each, from
which the posteriors were combined. Since \multi\ estimates the marginal
likelihood, we repeated our fits with $\eta$ fixed to zero and removed as a
free parameter, giving us a direct estimate of the Bayes factor, $B_{SP}$, for
the moon model. 

\begin{figure*}
\begin{center}
\includegraphics[width=18.0cm,angle=0,clip=true]{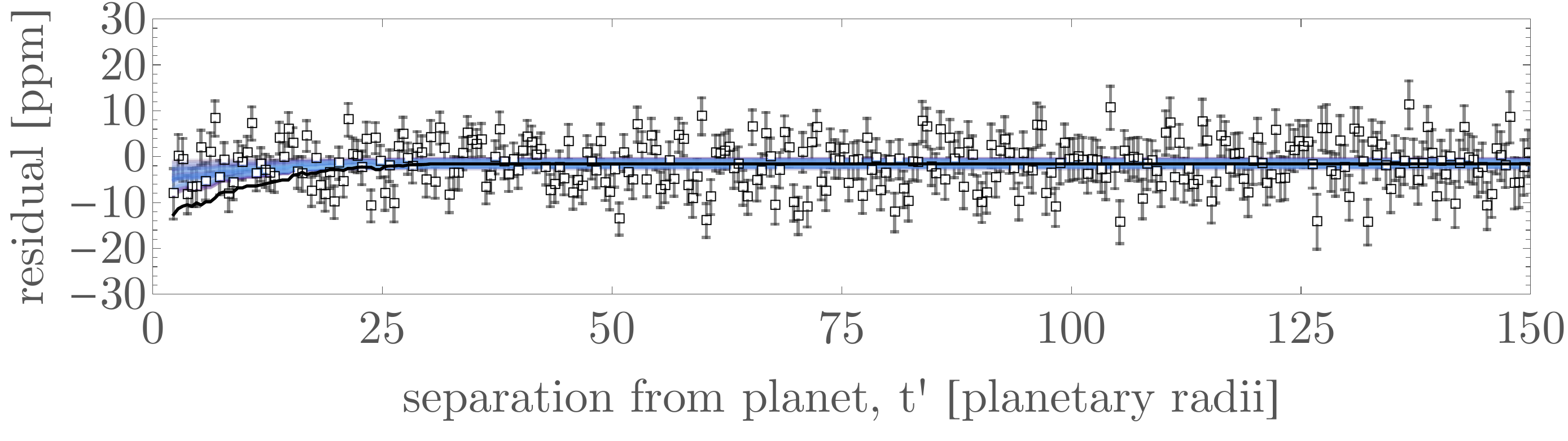}
\caption{
Phase-folded planet-stacked light curve of all 284 KOIs deemed to be of
acceptable quality. Temporal axis has been re-scaled and binned, with
uncertainties shown given by the standard deviations within each bin.
Black solid line represents the expected signature if $\eta=100$\% of the KOIs
had a Galilean-analog moon system. Blue lines show 100 posterior samples
from our fits, giving $\eta=0.16_{-0.10}^{+0.13}$.
}
\label{fig:ensemble}
\end{center}
\end{figure*}

\begin{figure}
\begin{center}
\includegraphics[width=8.4cm,angle=0,clip=true]{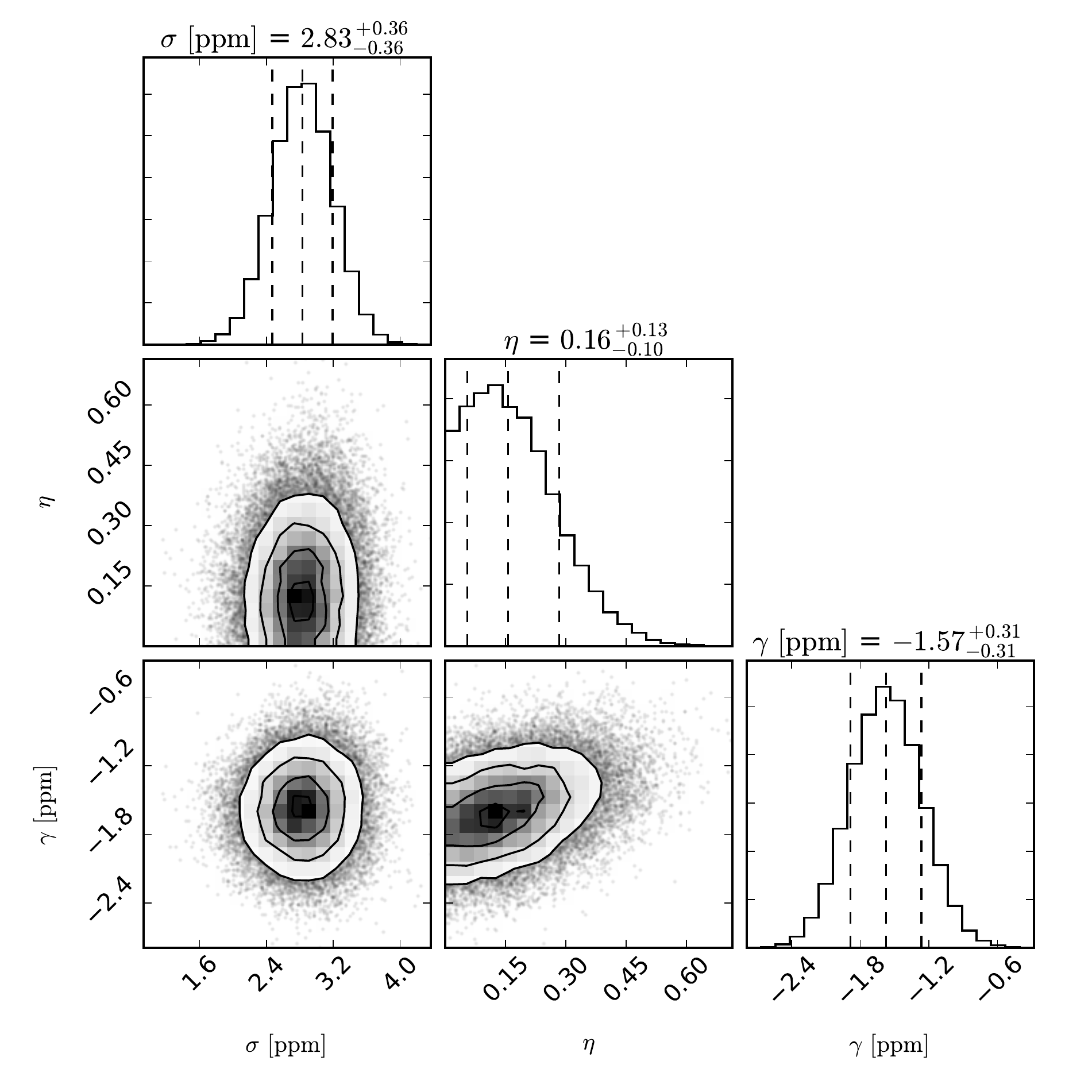}
\caption{
Corner plot of the three-parameter joint posterior distribution from
our Galilean-analog moon fit. This fit was disfavored over a null fit
with a Bayes factor of $B_{SP}=0.43$.
}
\label{fig:ensemble_post}
\end{center}
\end{figure}

We find that the null model is slightly favored, with $\log B_{SP} = -0.84$, or
an odds ratio of $2.3$-to-$1$ preference for the null model. The resulting
light curve and model fitting lines are shown in Figure~\ref{fig:ensemble},
and the associated posterior distribution is plotted in
Figure~\ref{fig:ensemble_post}. Our results imply that $\eta<0.38$ to 95\%
confidence for the 284 KOIs considered, with a 68.3\% confidence interval of $\eta=0.16_{-0.10}^{+0.13}$.

\subsection{Single Moon Global Fits}

As discussed in Section~\ref{subsub:single_moon_LUT}, the single moon case required
on-the-fly interpolation of a look-up table for the likelihood calls. For
this reason, we found it more practical to use a Markov Chain Monte Carlo (MCMC) algorithm
instead of \multi. Our regression was performed using a simple MCMC that we
wrote, which used the Metropolis rule for sampling and normal proposal
functions tuned by hand to give a ${\sim}50\%$ acceptance rate. Ten independent
chains were seeded from random locations within the prior volume, all of which
converged within 100 steps, and were then allowed to propagate for $10^5$
accepted steps, giving $\sim10^6$ posterior samples in total.

In addition to the three free parameters used in the Galilean fit, we added in
a parameter controlling the semi-major axis of the moon, $(a_{SP}/R_P)$, and
the moon size in Earth radii, $(R_S/R_{\oplus})$. Both were assumed to follow
log-uniform priors spanning the limits in our LUT. The $\eta$ term was allowed to
follow a log-uniform prior spanning 0.01 to 100, since technically it represents
an effective moon in this case and thus should be interpreted as the average number
of moons per system.

The fits converged to a solution of $R_S = 0.51_{-0.23}^{+0.59}$\,$R_{\oplus}$ and $(a_{SP}/R_P) = 6.3_{-3.1}^{+7.6}$ for $\eta = 0.43_{-0.28}^{+0.33}$. This fit does not directly return a marginal likelihood, since MCMC was used, nor is the Savage-Dickey ratio suitable given that three extra covariant free parameters have been added. However, the kernel-approach shown later reveals that the evidence favoring the single moon fit in this region is modest at $B_{SP}\simeq2$.

\subsection{Single Moon Kernel}

While the single moon fit is useful for identifying the maximum \textit{a posteriori} region of parameter space, it does not provide a clear view of the overall likelihood trends occurring within the prior volume. To address this, we repeated our fits for a single moon but fixed $a_{SP}/R_P$ and $R_S$ to a specific choice and just regressed $\eta$, $\sigma$ and $\gamma$. Since no interpolation was necessary, it was straightforward to use \multi\ with $\eta$-rescaling on a single interpolated model each time. $R_S$ was varied across a grid from $0.02$ to $2$ Earth radii in 100 log-evenly spaced steps. Similarly, $a_{SP}/R_P$ was varied from $2$ to $100$ in 100 log-even steps.

At each point, we derived a three-dimensional joint posterior distribution and marginalize over $\sigma$ and $\gamma$ to directly measure the occurrence rate of exomoons at each location. The posterior derived is mathematically equivalent to

\begin{align}
\mathrm{P}(\eta|R_S,a_{SP}/R_P) &= \int \int \mathrm{P}(\eta,\sigma,\gamma|R_S,a_{SP}/R_P) \mathrm{P}(\sigma)\mathrm{P}(\gamma)
\mathrm{d}\sigma \mathrm{d}\gamma
\end{align}

\begin{figure*}
\begin{center}
\includegraphics[width=18.0cm,angle=0,clip=true]{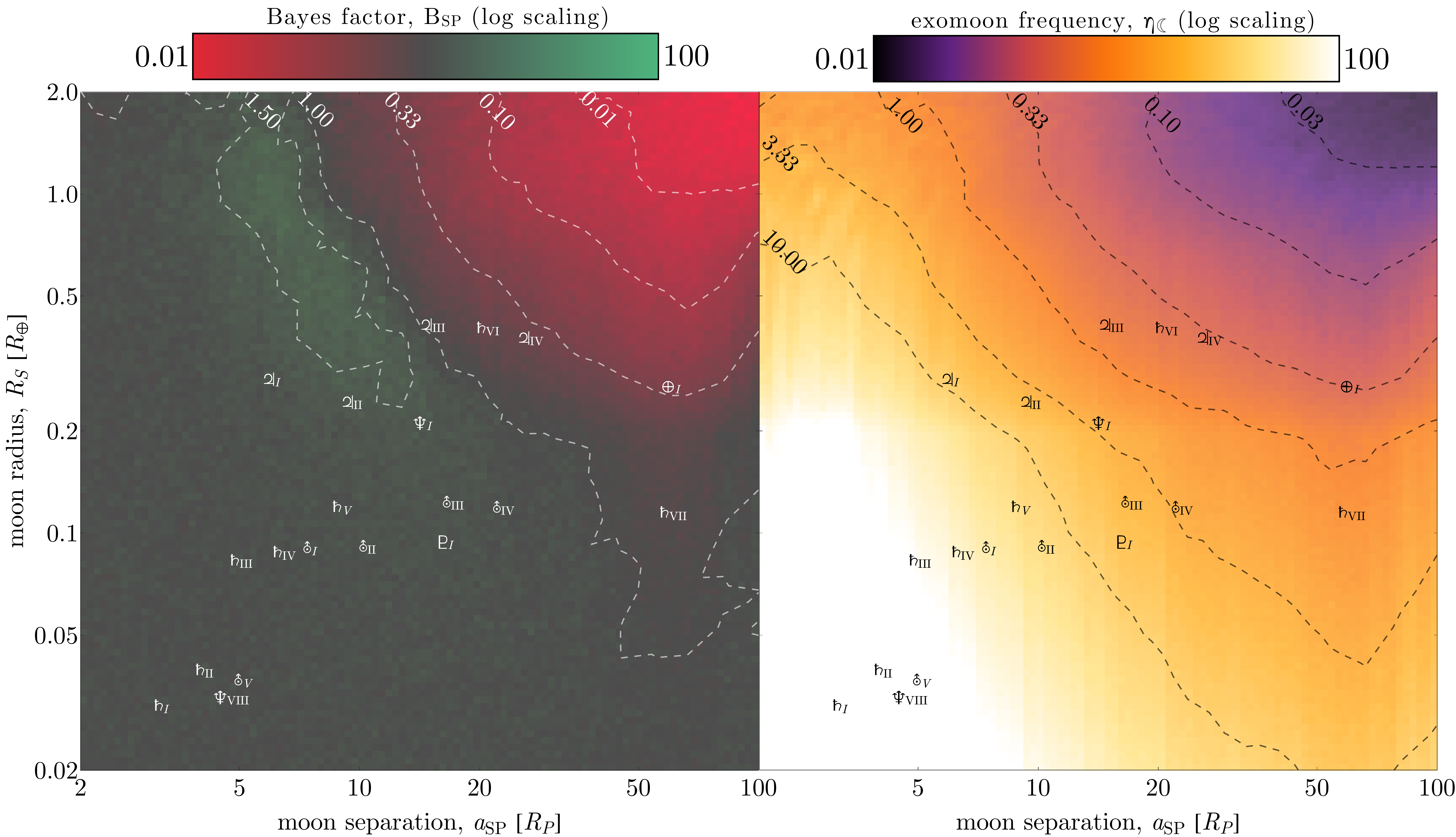}
\caption{
\textit{Left:} Heatmap of the Bayes factor $B_{SP}$ as a function of single effective moon radius $R_S$ and semi-major axis $a_{SP}$ for the ensemble. Red indicates regions of parameter space where the moon model is disfavored, while green represents regions where the moon model is favored. Greater color intensity corresponds to greater confidence in the selected model. \textit{Right:} Exomoon frequency in the ensemble as a function of $R_S$ and $a_{SP}$. A collection of solar system moons are plotted for context.
}
\label{fig:bayes_factor_and_moon_freq}
\end{center}
\end{figure*}

In addition to deriving a posterior at each grid point, we also estimate the evidence against the null model, allowing us to compute the Bayes factor.
In Figure \ref{fig:bayes_factor_and_moon_freq} we plot the Bayes factor and exomoon frequency for the ensemble as a function of effective moon radius $R_S$ and semi-major axis $a_{SP}$. 
The Bayes factor (left) indicates whether the moon model is favored over the model without a moon. 
Red represents regions of parameter space where the moon model is disfavored, while green regions are areas where the moon model is favored, and intensity represents our degree of confidence in that model selection. 
We emphasize to the reader that paying attention to the contours in this plot is essential for an accurate interpretation; while much of the plot appears green, the moon model is in fact only weakly favored ($B_{SP}\simeq2$) on a small island in parameter space.  
By contrast, large values of $R_S$ and $a_{SP}$ are strongly ruled out ($B_{SP}$ around 0.01). 
A value of 1 in this plot means we can make no statement about one model being a better fit to the data than the other.

The right side of Figure \ref{fig:bayes_factor_and_moon_freq} should then be read in the context of the left side. 
The similarity between the contours on both sides is readily apparent. 
For large values of $R_S$ and $a_{SP}$ we find an exceptionally low occurrence rate, while at the lower end of these variables the occurrence rate shoots up. It is tempting to read this as a moon signal, but in the context of the Bayes factor on left it is clear these occurrence rates are not at all constrained or well supported by the evidence. Only in regions of high confidence ($B_{SP}$ much greater or much less than 1) should the exomoon frequency values be given much credence.
It is worth pointing out, perhaps, that there is little qualitative difference between a very low exomoon occurrence rate and a very low value for $B_{SP}$.
Both are consistent with virtually no signal in this region of parameter space.

\subsection{Evidence for a Population of Super-Ios?}

The island on the left side of Figure \ref{fig:bayes_factor_and_moon_freq} where $B_{SP} > 1.50$ is intriguing, if only marginally significant. 
We hesitate to make any strong statement about this region of parameter space, but it is worth pointing out that theoretical modeling \citep[e.g.][]{namouni:2010} suggests that while planets migrating inward will tend to lose their moons in the process due to a shrinking Hill sphere, they are more likely to retain moons orbiting closer to the host planet (that is, close-in moons tend to survive planetary migration to smaller $a_P$). 
Recall that we are probing planets within roughly 1 AU of their host star, suggesting that a large fraction of these planets may have migrated from beyond the snow line.
The island of modest moon signal could therefore be evidence (albeit inconclusive) of a population of moons that have survived migration by virtue of their separation from their host planet.
Note, however, that more recent theoretical work \citep[][]{spalding:2016} suggests by contrast that satellites closer in to their planet ($a_{SP} \lesssim 10 R_P$) are also vulnerable to dynamical moon-loss during migration.
It is unclear, then, whether one or both of these mechanisms could be at play here, and indeed, the strength of these mechanisms rely in part of the evolutionary history of each system which will of course be unique.
In any case, our results suggesting a dearth of exomoons at small $a_P$ appear to provide observational support for the findings of both \citet{namouni:2010} and \citet{spalding:2016}, and more broadly, could be evidence of giant planet migration.

\subsection{Subset Fits}

\begin{figure*}[ht]

\includegraphics[angle=90, width=\textwidth]{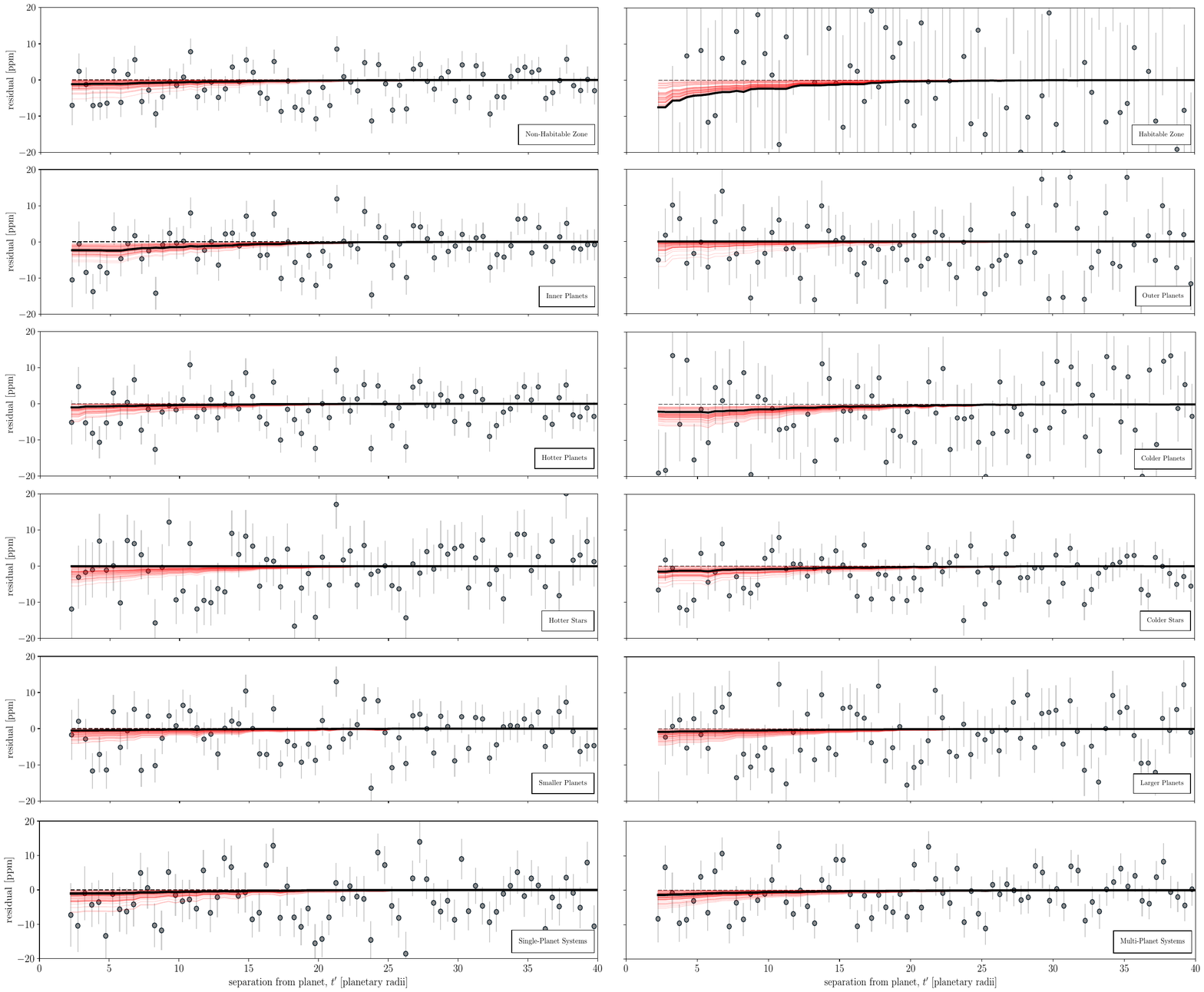}
\caption{Galilean analog GLC plots for a variety of sample subsets.}
\label{fig:GOSE_subsets}
\end{figure*}

\begin{table*}
\centering
\begin{tabular}{ | c | c | c| c || c | c | c | c |}
\hline
\multicolumn{8}{|c|}{Galilean Analog Subset Fits} \\
\hline
Group & $\eta_{\leftmoon}$\,[1\,$\sigma$] & $\eta_{\leftmoon}^{\mathrm{max}}$ [2\,$\sigma$] & $B_{SP}$ & Group & $\eta_{\leftmoon}$\,[1\,$\sigma$] & $\eta_{\leftmoon}^{\mathrm{max}}$ [2\,$\sigma$] & $B_{SP}$ \\
\hline
Smaller Planets & [0.06, 0.35] & 0.48 & $0.353\pm0.035$ &
Larger Planets & [0.07, 0.44] & 0.61 & $0.448\pm0.035$ \\
Colder Planets & [0.15, 0.68] & 0.86 & $1.003\pm0.070$ &
Hotter Planets & [0.07, 0.39] & 0.53 & $0.426\pm0.043$ \\
Colder Stars & [0.07, 0.33] & 0.44 & $0.639\pm0.063$ &
Hotter Stars & [0.06, 0.47] & 0.66 & $0.411\pm0.034$ \\
Inner Planets & [0.21, 0.64] & 0.80 & $2.564\pm0.250$ &
Outer Planets & [0.03, 0.28] & 0.42 & $0.224\pm0.018$ \\
Single-Planet Systems & [0.08, 0.50] & 0.68 & $0.689\pm0.066$ &
Multi-Planet Systems & [0.07, 0.33] & 0.44 & $0.420\pm0.038$ \\
Habitable Zone & [0.23, 0.88] & 0.97 & $1.679\pm0.083$ &
Non-habitable Zone & [0.08, 0.37] & 0.50 & $0.653\pm0.066$ \\
\hline
\end{tabular}
\caption{Table of occurrence rates $\eta$ for various subsets of the 284 planets examined in this work. Here $\eta$ represents 1$\sigma$ credible interval values from the posterior distributions while 95 pct is the 95\textsuperscript{th} percentile upper limit. $B_{SP}$ is the Bayesian evidence computed by \multi.}
\label{tab:galilean_fits}
\end{table*}

\begin{table*}
\centering
\begin{tabular}{ | c | c | c| c || c | c | c | c |}
\hline
\multicolumn{8}{|c|}{Single Effective Moon Subset Fits} \\
\hline
Group & $R_S$\,[1\,$\sigma$] & $R_S^{\mathrm{max}}$ [2\,$\sigma$] & $B_{SP}$ & Group & $R_S$\,[1\,$\sigma$] & $R_S^{\mathrm{max}}$ [2\,$\sigma$] & $B_{SP}$ \\
\hline
Smaller Planets & [0.02, 0.36] & 0.90 & 0.75 &
Larger Planets & [0.02, 0.41] & 1.18 & 0.76 \\
Colder Planets & [0.02, 0.41] & 1.12 & 0.81 &
Hotter Planets & [0.02, 0.42] & 1.03 & 0.79 \\
Colder Stars & [0.02, 0.35] & 0.83 & 0.79 &
Hotter Stars & [0.02, 0.36] & 1.07 & 0.74 \\
Inner Planets & [0.03, 0.81] & 1.40 & 1.04 &
Outer Planets & [0.02, 0.29] & 0.83 & 0.69 \\
Single-Planet Systems & [0.02, 0.42] & 1.05 & 0.79 &
Multi-Planet Systems & [0.02, 0.34] & 0.96 & 0.72 \\
Habitable Zone & [0.03, 1.10] & 1.66 & 1.12 &
Non-habitable Zone & [0.02, 0.33] & 0.92 & 0.74 \\
\hline
\end{tabular}
\caption{Table of effective moon sizes $R_S$ for various subsets of the 284 planets examined in this work, in units of Earth radii. We present 1$\sigma$ credible interval values from the posterior distributions while 95 pct is the 95\textsuperscript{th} percentile upper limit. Here $B_{SP}$ is the Savage-Dickey ratio computed from the $\log(R_S)$ posteriors.}
\label{table:effective_moon_subsets}
\end{table*}

In addition to fitting the entire sample with effective moon and Galilean analog GLC models, we also performed GLC fits on a number of physically motivated subsets. 
The aim here was to identify whether a certain class of planets in the sample preferentially hosts moons over another. 
As such we divided the sample into several equally-populated pairs and fit the GLC model to these subsets. 
These pairs were small/large planets, cold/hot planets, cold/hot stars, and inner/outer planets.
We also split the sample into single/multi-planet systems, and habitable zone/non-habitable zone planets (most of the latter residing inside the innermost edge of the habitable zone). 
These last two categories, of course, are not equally populated. Insolations were taken from \NEA\, and anything less than the maximum (inner-edge) insolation given in \citet{yang:2014} equation 2 
was considered to be in the habitable zone.

The results for Galilean moon fits can be seen in Figure \ref{fig:GOSE_subsets} and Table \ref{tab:galilean_fits}. The thick black line in the figure 
represents the peak posterior value, while the light red lines represent 50 fair draws from the posterior.
From these plots we can make a number of comparisons.
Dividing the sample in two by size, we see a marginally higher occurrence rate for the larger planets.
For planet temperature we see a higher occurrence rate on the colder end.
We also see a higher exomoon occurrence rate for colder stars, which we can take to mean later-type or evolved stars.
All of these observations are in line with what we might expect \textit{a priori}. 

We see very little difference in the occurrence rate for single- and multi-planet systems. 
Interestingly, though, the inner 50\% of planets (those closest to their star) show a significantly higher exomoon occurrence rate than those farther away, where the maximum \textit{a posteriori} value indicates a total non-detection. 
This is unexpected, since the Hill sphere shrinks with smaller semi-major axis.
Finally, and perhaps most intriguingly, the maximum \textit{a posteriori} values for habitable zone planets ran away to the maximum, while non-habitable zone planets have a much lower occurrence rate.
This should be read with caution, however, considering the size of the error bars in the habitable zone planet case. 
While the comparisons above are made between equally populated subsets, there are far more non-habitable zone planets than there are planets in the habitable zone, making the results in the latter case far more uncertain.

Single, effective moon fits were also performed for these same OSE subsets. Results can be found in Table \ref{table:effective_moon_subsets}.
Unlike the case for Galilean analog fits, we cannot meaningfully quote an occurrence rate in this case because the depth of the moon signal is controlled by the size of the effective moon.
There is a degeneracy between moon size and occurrence rate, so we model the size of the moon as a proxy for occurrence rate.
In essence, a smaller effective moon can mean either a) we have a high occurrence rate, with small moons, or b) we have a low occurrence rate of larger moons. 
The truth, of course, is probably somewhere in the middle.

To characterize whether our fits exclude a null detection to high confidence we compute the Savage-Dickey ratio between a uniform prior and the posterior distribution for $R_S^2 = 0$. We treat this number as our Bayes factor. For the colder stars we find $B_{SP} = 1.53$, for outer planets $B_{SP} = 0.7$, large planets $B_{SP} = 1.29$, multi-planet systems $B_{SP} = 0.77$, and habitable zone planets $B_{SP} = 3.11$.
These values suggest that there is only marginal evidence in support of an effective moon signal in the cold star, large planet, and habitable zone planet subsets, while the null hypothesis is favored for the outer planet and multi-planet subsets.

\section{EXOMOON CANDIDATE KEPLER-1625\lowercase{b} I}
\label{sec:kepler1625b}

\subsection{Individual Fits}

Thus far, our analysis has taken a population-based approach to seeking exomoons,
in contrast to the conventional method adopted in previous HEK papers where candidates
were interrogated individually. Although a full suite of photodynamical Bayesian
fits to each planetary candidate is beyond the scope of this work, representing
a formidable computational challenge, we do here investigate systems individually
with a simpler model.

For each KOI, we took the final phase-folded light curve and used \multi\ to
fit the \citet{heller:2014} OSE model through the data. As discussed earlier,
OSE does have drawbacks, yet it remains a useful and quick tool for checking
for any significant flux decreases surrounding the phase-folded transits. For
each KOI, we fit for an offset term, a photometric jitter, the moon size, and a
semi-major axis, $(a_{SP}/R_P)$ with \multi\ using 1000 live points. In the
earlier ensemble fits, we used a log-uniform prior on the moon-size (see Section~\ref{sub:galileanfits}), which means negative moon radii cannot be
sampled. In these fits, we wished to allow for negative radius moon, which
correspond to inverted transit signals, to provide insights into possible
biases affecting the results. Accordingly, we modified our moon parameter to
be the satellite-to-star ratio-of-radii, $s$, squared, to account for the
fact that transit detection bias is approximately proportional to $s^2$.

\begin{figure}
    \centering
    \includegraphics[width=8.4cm]{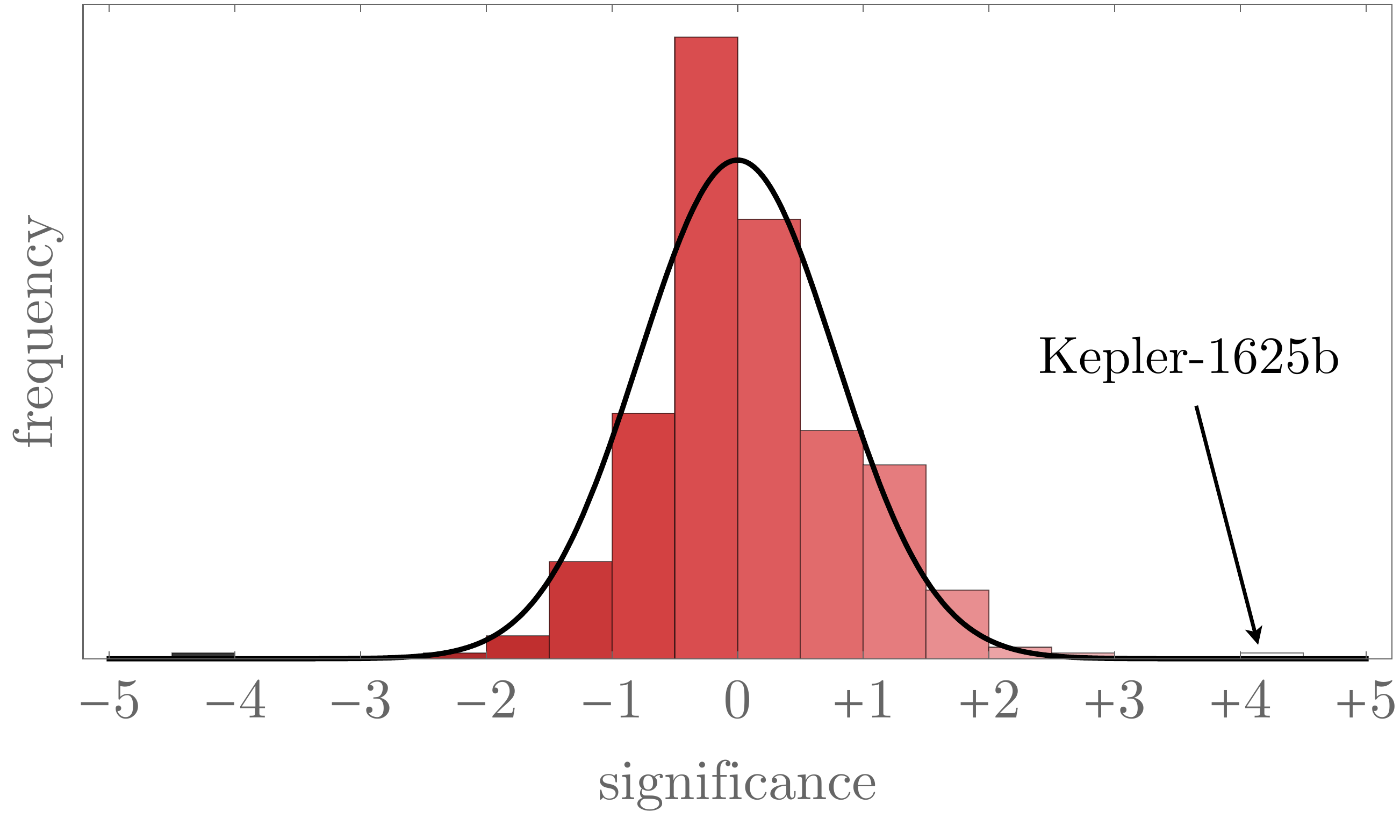}
    \caption{Histogram of the ``significance'' of an OSE detection for several hundred KOIs, the test which revealed the presence of a possible candidate around Kepler-1625b. The vertical axis scale is linear.}
    \label{fig:osehisto}
\end{figure}

We evaluated the median $s^2$ value for each KOI from the posterior chain
and divided by the lower quantile bounding a 68.3\% confidence interval, a metric
which we loosely refer to as ``significance'' in what follows (although it
is best interpreted as a power). In Figure~\ref{fig:osehisto}, we show a histogram
of the resulting significances for 353 KOIs, where we have also included the
likely eccentric KOIs to provide a wider sample to assess distribution properties.
Inspection of the results reveals a sizable spread centered around zero, as
might be expected. However, we note that KOI-5084.01 (Kepler-1625b) appears quite
deviant from the bulk population with a $+4.4$\,$\sigma$ significance. Similarly,
KOI-4202.01 (Kepler-1567b) is an outlier on the negative scale with a
$-4.4$\,$\sigma$ significance.

The negative outlier clearly cannot be a genuine exomoon and thus we do not consider
it further it in what follows. The positive outlier though could be an isolated
candidate missed by the ensemble analyses. Excluding the two outliers, the histogram
shown in Figure~\ref{fig:osehisto} is well-described by a normal distribution
centered on zero with a standard deviation of 0.8. If we generate a list of 353
random variates from that distribution and take the maximum value, $+4.4$ is
highly improbable; we used Monte Carlo simulations to estimate the
probability, which was found to be $4 \times 10^{-6}$. On this basis, Kepler-1625b appears quite
unexpected and thus worthy of more detailed investigation.

\subsection{Detailed Investigation of Kepler-1625b}

To investigate further, we performed detailed photodynamical fits of Kepler-1625b
using the \luna\ model and \multi, in the same manner as that conducted in our
previous series of HEK papers (e.g. see \citealt{HEK3}). This enables a rigorous
Bayesian model selection to ensure not only a physically plausible model can
explain the photometry, but that the moon parameters are justified given the
extra complexity they introduce.

\begin{figure}
    \centering
    \includegraphics[width=8.4cm]{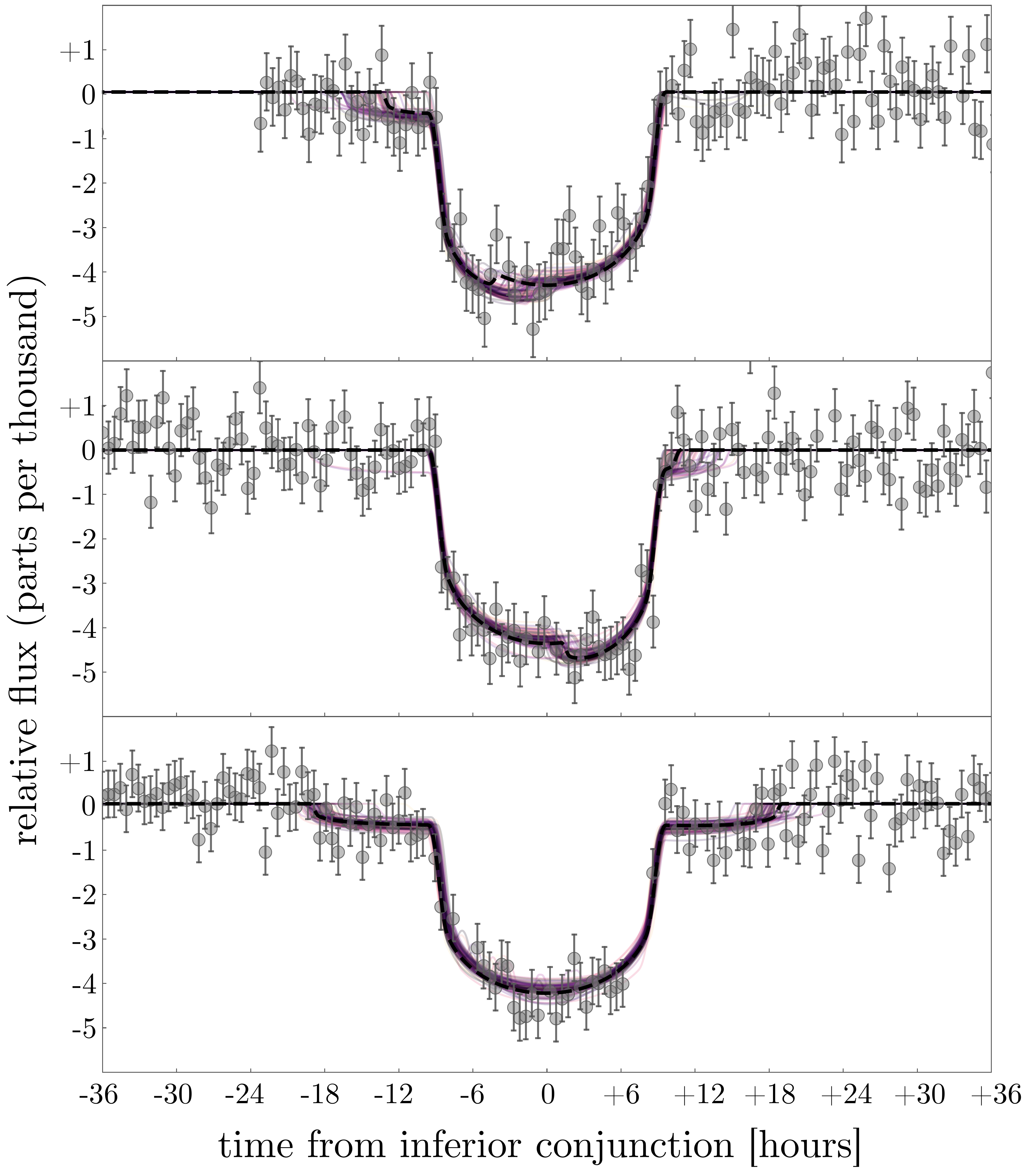}
    \caption{The three transits of Kepler-1625b observed with \textit{Kepler}, overlaid with 100 draws from the model posteriors. The black line is the maximum \textit{a posteriori} model.
    }
    \label{fig:kepler1625_transits}
\end{figure}

Comparing the evidences between a planet-only and planet+moon model revealed
that the moon model was favored with $\log B_{SP} = 10.2$, or a 4.1\,$\sigma$
preference, consistent with the level found previously. The light curve fits
are illustrated in Figure~\ref{fig:kepler1625_transits}.

Given the limited numbed of transits available, just three, the conventional
HEK approach of cross-validating by removing a transit was not tenable and
thus we adapted this strategy somewhat in what follows. Instead, we performed
$k$-fold cross-validations, where we omit one-half of a transit (centered about
the time of inferior conjunction) and re-fit the remaining data blindly each
time.  Since there are three transits in the \textit{Kepler} data, this yields
six unique ways of conducting the cross-validation. In each case, we performed
a new blind fit and found a positive detection consistent with the original
signal in all cases, with Bayes factors indicating 3.9, 2.4, 4.6,
4.0, 3.1, and 2.7 $\sigma$ preference for the moon model (comparable to the
original 4.1\,$\sigma$ detection when using all of the data).

As with all previous moon candidates we attempted to rule out all other possible
explanations for the signal. For any candidate moon signal there will be one of
two possible explanations: the signal is either an instrumental artifact or it
has a true astrophysical origin (be it an exomoon or something else).  To test
the possibility of an instrumental aberration we performed an independent and
manual detrending on the \textit{Kepler} DR25 data (our OSE survey used DR24)
using \cofiam\ on the PA and PDC data. In both cases the planet-moon model
remained the favored hypothesis over the planet-only model. In contrast,
polynomial-based detrending was found to remove the signal, likely due to very
long-timescale nature of the driving event occurring in epoch three.

We also examined data from individual \textit{Kepler} pixels to determine whether
there might have been anomalous behavior in the vicinity of the transit events
(as occurred for false-positive Kepler-90g.01; \citealt{kepler90}), but no unusual
spikes or drop outs were detected in any relevant pixel. We also verified that there
were no bad data flags at the time of the transits, which was a source of
the false-positive result for the moon candidate around PH2-b \citep{HEK5}.

If the signal were astrophysical in origin, then there are several possible hypotheses, including a moon. A ring is not likely since it should produce a coherent signal in
all events, which is not seen, unless the ring precession rate is very fast. 
Rotating spots on the surface of the star also affect the light curve, primarily
producing a long-term undulation in the data. This long-term trend is removed via our detrending (repeated several times independently) and thus the only remaining possible starspot-induced signal would be crossing events. However, starspot crossings (when a
transiting planet occults a dark spot; e.g. see \citealt{rabus:2009}) cannot be
responsible for the observed moon-like dips. This is because spot crossings can only
occur inside the main planetary transit and can never produce out-of-transit flux
decreases, as seen for Kepler-1625b, purely from a geometrical argument. 
If the signal were confirmed then, this would leave the exomoon hypothesis as the
leading explanation based on current information.

The quoted stellar properties of Kepler-1625 in \NEA\, changed significantly from DR24 to DR25, owing to the addition of updated information in the latter data release \citep{mathur:2017}. This update pushed the star from a sub-Solar to a
super-Solar radius ($R_{\star} = 0.838^{+0.366}_{-0.079} \rightarrow 1.793^{+0.263}_{-0.488} \: R_{\astrosun}$), enhanced the metallicity from sub- to super-Solar abundances, and lowered the density substantially ($\rho_{\star} = 2.059^{+0.4626}_{-1.306} \rightarrow 0.2636^{+0.3257}_{-0.0768} \:$ g/cm$^3$), indicating that this star is likely climbing the giant branch. Critically, our planet-only and planet+moon fits favor a low stellar density
of $\rho_{\star} = 0.387_{-0.083}^{+0.034}$\,g/cm$^3$ and $\rho_{\star} = 0.405_{-0.054}^{+0.028}$\,g/cm$^3$ and if the true density were much higher, then Kepler-1625b
would need to be either highly eccentric or blended \citep{AP:2014}, both of which would be severely detrimental to the exomoon hypothesis. Determining the true nature of this star is critical as it will also dictate the sizes of the planet and moon derived from the transit depth (which we describe shortly).

We also attempted to recover a rotation period for the star (following the methodology described in \citealt{torres:2015}) but the amplitude of variability appears too small to recover a consistent period across each quarter, with best-fitting periods ranging from 4.5 days to 21\,days. Attempting to regress a coherent signal across all quarters gives an amplitude of 66\,ppm, and when performed on each quarter independently, the median amplitude was 136\,ppm. Given the lack of strong evidence for rotation, the weak amplitudes in comparison to the candidate moon transit depth (570\,ppm), and the arguments made earlier as to why rotational modulations are unlikely to be a source of false-positive, we deem it unlikely that activity is responsible for the signal observed.

Our photodynamical fits combined with the DR25 stellar properties indicate that Kepler-1625b is likely a Jupiter-sized planet with approximately ten times Jupiter's mass, orbited by a moon roughly the size of Neptune. We calculate the radii of the planet and moon by measuring the depth of the flux dip ($\Delta F / F = (R_o / R_{*})^2$, where $R_o$ is the radius of the object in question) and we are able to derive a mass based on the photodynamical model fit.
We note that both the planet and the moon show good agreement between mass and radius estimates and lie in physically reasonable parameter space based on the mass-radius forecaster model of \citet{chen:2017}.
We find the semi-major axis of the moon $a_S = 19.1_{-1.9}^{+2.1} \: R_P$, which is well outside the Roche limit and comfortably within the Hill sphere for this planet. 
It is dynamically stable and should not have spun out / escaped over 5 Gyr \citep{barnes:2002}.

While the existence of a Neptune-sized moon has largely not been anticipated in the literature (however, see \citealt{cabrera:2007}), we cannot readily rule out its existence on these grounds. 
Indeed, the existence of Hot-Jupiters was also wholly unexpected prior to their discovery in the mid-1990s. 
It seems clear that a moon of this type could not have formed in a circumplanetary accretion disk akin to that which is thought to have formed the regular moons of Jupiter and Saturn. 
It is conceivable however that the moon could have been captured by the planet, a kind of intermediate process between typical capture scenarios (e.g. Neptune and Triton; see \citealt{agnor:2006}) and the cataclysmic impact event that is believed to have formed Earth's Moon \citep{cuk:2012}. 
In this scenario a grazing impact might be experienced as a kind of extreme atmospheric drag sufficient to capture the passing body. 
Observation of this system might therefore not only produce the first unambiguous detection of an exomoon, but could also go a long way in demonstrating once again that what we observe in our Solar System is not all that is possible. 

\subsection{Validating the Exomoon Candidate Kepler-1625b I}

At this time, we remain cautious about the reality of this signal, given the relatively small number of transits available. This is particularly true because the third transit appears to be crucial to the exomoon interpretation and can be removed using polynomial-based detrending approaches. Detrending the photometric time series of long-period transits is more challenging than their shorter-period counterparts and it remains wholly plausible that the signal observed is nothing more than an artifact of our detrending process. We strongly emphasize these points and encourage the community to not treat this signal as genuine until it can be confirmed.

Fortunately, our photodynamic moon fit yields a testable prediction for the morphology of the next transit event occurring October 2017. With such a long event duration from anticipated exomoon ingress to egress, the event cannot be observed in its entirety by any single optical/NIR instrument on the ground south of latitude $\sim 78^{\circ}$ N (north of which the long Arctic night has already begun on the date of observation). A space-based observation is clearly essential to characterize the system. We have therefore secured HST observations to validate the signal during the next transit of the planet, and we strongly advocate treating this object as no more than a candidate at this time, similar to previous moon candidates discussed in earlier HEK papers.



\section{CONCLUSION}
\label{sec:conclusion}

In this work we have examined 284 \textit{Kepler} exoplanets (from an original sample of 4098 KOIs) in search
of an exomoon signal in the ensemble. We performed a rigorous multi-stage analysis to select only the
highest quality data, measure and correct for TTVs, and stack a total of 6096 transit events to characterize the exomoon population. 
As a byproduct of our work we present new TTV posterior distributions, along with a handful of stellar properties, and make
them available online to the community. 

Our results place new upper limits on the exomoon population for 
planets orbiting within about 1 AU of their host star, upper limits that are remarkably low. 
We have also analyzed subsets of the ensemble to test the effect of various data cuts. 
Our analysis suggests that exomoons may be quite rare 
around planets at small semi-major axes, a finding that supports theoretical work suggesting moons may
be lost as planets migrate inward. On the other hand, if the dearth of exomoons can be read as a reliable
indicator of migration, our results suggest a large fraction of the planets in the ensemble have
migrated to their present location. 

Finally, we have briefly highlighted our identification of an exomoon candidate in the Kepler-1625 system, for which we have secured a follow-up observation with HST.
This candidate has passed a thorough preliminary inspection, but we emphasize again our position that the \textit{Kepler} data are insufficient to make a conclusive
statement about the existence of this moon. Only after the HST observation is made should any claim about this moon's existence be given much credence.

\vspace{2cm}
\acknowledgments

This paper includes data collected by the \emph{Kepler} mission. Funding for the 
\emph{Kepler} mission is provided by the NASA Science Mission directorate.

Resources supporting this work were provided by the NASA High-End Computing
(HEC) Program through the NASA Advanced Supercomputing (NAS) Division at Ames 
Research Center.

This research has made use of the NASA Exoplanet Archive, which is operated by
the California Institute of Technology, under contract with the National 
Aeronautics and Space Administration under the Exoplanet Exploration Program.

This research has made use of the {\tt corner.py} code by Dan Foreman-Mackey at 
\href{http://github.com/dfm/corner.py}{github.com/dfm/corner.py}.

This work made use of the Michael Dodds Computing Facility, for which
we are grateful to Michael Dodds, 
Carl Allegretti, David Van Buren, Anthony Grange, Cameron Lehman, Ivan Longland, 
Dell Lunceford, Gregor Rothfuss, Matt Salzberg, Richard Sundvall, Graham 
Symmonds, Kenneth Takigawa,
Marion Adam, Dour High Arch, Mike Barrett, Greg Cole, Sheena Dean, Steven 
Delong, Robert Goodman, Mark Greene, Stephen Kitt, Robert Leyland, Matthias 
Meier, Roy Mitsuoka, David Nicholson, Nicole Papas, Steven Purcell, Austen 
Redman, Michael Sheldon, Ronald Sonenthal, Nicholas Steinbrecher, Corbin Sydney, 
John Vajgrt, Louise Valmoria, Hunter Williams, Troy Winarski and Nigel Wright.

We thank members of the Cool Worlds Lab for helpful conversations in preparing this manuscript.
Finally, we thank the anonymous referees for their constructive comments.
DMK acknowledges support from NASA grant NNX15AF09G (NASA ADAP Program). AT acknowledges support from the NSF GRFP grant DGE 16-44869.

\end{document}